\documentclass[12pt,amssymb,epsf]{article}
\usepackage{epsfig}
\usepackage{axodraw}
\newcommand{\li}{\mbox{${\cal L}{\mbox i}_{2}$}} 
\newcommand{\Li}{\mbox{${\mbox L}{\mbox i}_{2}$}}

\newcommand{\real}{{\cal\mbox{Re\,}}}
\newcommand{\imag}{{\cal\mbox{Im\,}}}

\newcommand{\pole}{\mbox{${\cal P}ole$}}
\newcommand{\ff}{\mbox{${\cal F}$}}
\newcommand{\R}{\mbox{${\cal R}$}}
\newcommand{\be}{\begin{equation}}
\newcommand{\ee}{\end{equation}}
\newcommand{\pvec}{\mbox{$\vec{\, p}$}}

\begin{document}
\pagestyle{empty}
\begin{flushright}
{CERN-TH/97--158}
\end{flushright}
\vspace*{5mm}
\begin{center}
    {\bf NON-FACTORIZABLE CORRECTIONS TO W-PAIR PRODUCTION: \\ 
METHODS AND ANALYTIC RESULTS } \\
\vspace*{1cm} 
{\bf W.~Beenakker}$^{*)}$,\ \ {\bf A.P.~Chapovsky$^{\dagger)}$}\\
\vspace{0.3cm}
Instituut--Lorentz, University of Leiden, The Netherlands 
\vspace{0.6cm}\\
and 
\vspace{0.5cm}\\
{\bf F.A.~Berends} 
\vspace{0.3cm}\\
Theory Division, CERN, CH--1211 Geneva 23, Switzerland 
\vspace{0.3cm}\\ 
and 
\vspace{0.3cm}\\ 
Instituut--Lorentz, University of Leiden, The Netherlands \\
\vspace*{2cm}  
                                {\bf ABSTRACT} \\ \end{center}
\vspace*{5mm}
\noindent
In this paper we present two methods to evaluate non-factorizable corrections 
to pair-production of unstable particles. The methods are illustrated in detail
for $W$-pair-mediated four-fermion production. The results are valid a few 
widths above threshold, but not at threshold. One method uses the decomposition
of $n$-point scalar functions for virtual and real photons, and can therefore 
be generalized to more complicated final states than four fermions. The other 
technique is an elaboration on a method known from the literature and serves 
as a useful check. Applications to other processes than $W$-pair production 
are briefly mentioned.
 \vspace*{1.2cm}\\ 
\begin{flushleft} CERN-TH/97--158 \\
July 1997
\end{flushleft}
\noindent 
\rule[.1in]{16.5cm}{.002in}

\noindent
$^{*)}$Research supported by a fellowship of the Royal Dutch Academy of Arts 
and Sciences.\\
$^{\dagger)}$Research supported by the Stichting FOM.
\vspace*{0.3cm}

\vfill\eject

\setcounter{page}{1}
\pagestyle{plain}


\section{Introduction}
\label{sec:intro}

With the start of LEP2, quantitative knowledge of the radiative corrections to
the four-fermion production process $e^+ e^- \to 4 f$
is needed \cite{review}. The full calculation of all these corrections will be
extremely involved and at present one relies on approximations \cite{review}, 
such as leading-log initial-state radiation and running couplings \cite{bhf}. 
Another approach is to exploit the fact that the corrections, in particular 
those associated with the production of an intermediate $W$-boson pair, are 
important.
This (charged-current) production mechanism dominates at LEP2 energies and 
determines the LEP2 sensitivity to the mass of the $W$ boson and to the 
non-Abelian triple gauge-boson interactions. As such, one could approximate the
complete set of radiative corrections by considering only the leading terms in
an expansion around the $W$ poles. The double-pole residues thus obtained 
could be viewed as a gauge-invariant definition of corrections to ``$W$-pair 
production''. The sub-leading terms in this expansion are generically 
suppressed by powers of $\Gamma_W/M_W$, with $M_W$ and $\Gamma_W$ denoting the 
mass and width of the $W$ boson. The quality of this double-pole approximation 
degrades in the vicinity of the $W$-pair production threshold, but a few 
$\Gamma_W$ above threshold it is already quite reliable \cite{pole-scheme}. 
It is conceivable that in the near future a combination of the above-mentioned 
approximations will result in sufficiently accurate theoretical predictions 
for four-fermion production processes.

In the double-pole approximation the complete set of first-order radiative 
corrections to the charged-current four-fermion processes can be divided into 
so-called factorizable and non-factorizable corrections 
\cite{review,pole-scheme}, i.e.~corrections that manifestly contain two 
resonant $W$ propagators as overall factors and those that do not.  In view of 
gauge-invariance requirements, some care has to be taken with the precise 
definition of this split-up (see below). In the factorizable corrections 
one can distinguish between corrections to $W$-pair production and $W$ decay. 
In this paper we give a detailed account of the non-factorizable corrections 
that were used in the analysis of \cite{nfletter}. From 
the complete set of electroweak Feynman 
diagrams that contribute to the full $\cal{O}(\alpha)$ correction, we will 
therefore only consider the non-factorizable ones, both for virtual corrections
and real-photon bremsstrahlung. To be more precise, since we are only 
interested in the double-pole terms we are led to consider only  
non-factorizable QED diagrams in the soft-photon limit. 
This means that we use simplified expressions for loop corrections and
real-bremsstrahlung interferences, i.e.~the photon momentum $k^{\mu}$ is
neglected in the numerators and whenever possible $k^{2}$ is neglected in the 
denominators. It does not mean that the photon energy is taken to be small in
the actual loop/phase-space integrations. In fact, for real bremsstrahlung
the photon is treated inclusively and the energy is extended to infinity, 
which simplifies and approximates the phase-space integrals \cite{fadin-khoze}.
The errors associated with this procedure are formally of higher order in the 
expansion in powers of $\Gamma_W/M_W$ in the energy region where 
$\Delta E\approx M_{W}$, 
since photons with an energy of ${\cal O}(M_{W})$ force the $W$-bosons off the 
resonance \cite{fadin-khoze}. Here $\Delta E$ stands  for the 
distance in energy to the threshold. In the energy region where 
$\Gamma_{W}\ll \Delta E\ll M_{W}$, the accuracy of
the approximation becomes of ${\cal O}(\Gamma_{W}/\Delta E)$.
Finally, near threshold, where $\Delta E\approx\Gamma_{W}$, our approximation 
breaks down. There the dominant
correction comes from the Coulomb effect, which was discussed in great detail
in the literature (see e.g. \cite{coulomb1,coulomb2,coulomb3}).

The above approach is the same as the one adopted by the authors of 
Ref.\,\cite{melyak}, 
who were the first to calculate non-factorizable $W$-pair corrections. For the 
present calculations, we have used two different methods. One is an extension 
of the treatment in \cite{melyak}, the other is a modification of the standard 
methods, which involves a combination of the decomposition of multipoint scalar
functions and the Feynman-parameter technique. The results obtained with our 
two methods are in complete mutual agreement. However, in contrast to 
\cite{melyak}, a clear separation between virtual and real photonic
corrections has been made in both methods, which is essential to establish the 
cancellations of infrared and collinear divergences. This treatment reveals 
a significant difference between our results and those obtained by the
authors of \cite{melyak}.
Our final results do not contain any logarithmic terms involving the 
final-state fermion masses, whereas in the results of \cite{melyak} explicit 
logarithms of fermion-mass ratios occur (see discussion in Sect.\,4.1 of 
Ref.\,\cite{melyak}). This difference can be traced back to
the fact that although the fermion masses can formally be neglected in the
absence of collinear divergences, they have to be introduced in intermediate 
results in order to regularize those divergences before dropping out from the 
final results.

\subsection{Gauge-invariant definition of non-factorizable corrections}
\label{sec:gaugeinv}

\begin{figure}
  \unitlength 0.95pt
  \begin{center}
  \begin{picture}(430,150)(0,0)
    \ArrowLine(30,75)(0,125)
    \ArrowLine(0,25)(30,75)
    \Photon(30,75)(80,110){1}{7} 
    \Photon(30,75)(80,40){1}{7}  
    \ArrowLine(120,90)(110,95) 
    \ArrowLine(110,95)(80,110) 
    \Photon(65,50.5)(110,95){1}{7}           
    \ArrowLine(80,110)(120,130)      
    \ArrowLine(120,20)(80,40)      
    \ArrowLine(80,40)(120,60)      
    \GCirc(30,75){15}{1}
    \GCirc(80,110){5}{1} 
    \GCirc(80,40){5}{1}  
    \Text(55,0)[]{\boldmath $D_{0123}$}
    \Text(10,123)[lb]{$e^+$}
    \Text(10,38)[lt]{$e^-$}
    \Text(55,105)[b]{$W^+$}
    \Text(53,53)[t]{$W^-$}
    \Text(100,76)[lt]{$\gamma$}
    \Text(130,137)[lb]{$k_1^{\prime}$}
    \Text(130,96)[lt]{$k_1$}
    \Text(130,64)[lb]{$k_2$}
    \Text(130,25)[lt]{$k_2^{\prime}$}
    \Text(89,79)[rb]{\boldmath $\scriptstyle k$}
    \Text(53,70)[lb]{\boldmath $\scriptstyle p_2+k$}
    \Text(64,100)[lt]{\boldmath $\scriptstyle p_1-k$}
    \Text(66,39)[lb]{\boldmath $\scriptstyle p_2$}
    \ArrowLine(170,75)(140,125)
    \ArrowLine(140,25)(170,75)
    \Photon(170,75)(220,110){1}{7} 
    \Photon(170,75)(220,40){1}{7}  
    \ArrowLine(260,90)(220,110)      
    \ArrowLine(220,110)(260,130)      
    \ArrowLine(260,20)(220,40)      
    \ArrowLine(220,40)(250,55)
    \ArrowLine(250,55)(260,60) 
    \Photon(205,99.5)(250,55){1}{7}     
    \GCirc(170,75){15}{1}
    \GCirc(220,110){5}{1} 
    \GCirc(220,40){5}{1} 
    \Text(204,0)[]{\boldmath $D_{0124}$} 
    \Text(202,104)[b]{$W^+$}
    \Text(203,53)[t]{$W^-$}
    \Text(249,87)[lt]{$\gamma$}
    \Text(279,137)[lb]{$k_1^{\prime}$}
    \Text(279,96)[lt]{$k_1$}
    \Text(279,64)[lb]{$k_2$}
    \Text(279,25)[lt]{$k_2^{\prime}$}
    \Text(237,82)[rt]{\boldmath $\scriptstyle k$}
    \Text(213,62)[lb]{\boldmath $\scriptstyle p_2+k$}
    \Text(202,93)[lt]{\boldmath $\scriptstyle p_1-k$}
    \Text(214,118)[lb]{\boldmath $\scriptstyle p_1$}
    \ArrowLine(310,75)(280,125)
    \ArrowLine(280,25)(310,75)
    \Photon(310,75)(360,110){1}{7} 
    \Photon(310,75)(360,40){1}{7}  
    \ArrowLine(400,90)(380,100)    
    \ArrowLine(380,100)(360,110)   
    \ArrowLine(360,110)(400,130)      
    \ArrowLine(400,20)(360,40)      
    \ArrowLine(360,40)(380,50)   
    \ArrowLine(380,50)(400,60) 
    \Photon(380,50)(380,100){1}{7}      
    \GCirc(310,75){15}{1}
    \GCirc(360,110){5}{1} 
    \GCirc(360,40){5}{1}
    \Text(355,0)[]{\boldmath $E_{01234}$}
    \Text(351,104)[b]{$W^+$}
    \Text(352,53)[t]{$W^-$}
    \Text(409,77)[l]{$\gamma$}
    \Text(427,137)[lb]{$k_1^{\prime}$}
    \Text(427,96)[lt]{$k_1$}
    \Text(427,64)[lb]{$k_2$}
    \Text(427,25)[lt]{$k_2^{\prime}$} 
    \Text(361,100)[lt]{\boldmath $\scriptstyle p_1-k$} 
    \Text(361,62)[lb]{\boldmath $\scriptstyle p_2+k$}
    \Text(394,80)[r]{\boldmath $\scriptstyle k$}
  \end{picture} 
  \caption[]{Virtual diagrams contributing to the manifestly non-factorizable 
             $W$-pair corrections in the purely leptonic case. The scalar 
             functions corresponding to these diagrams are denoted by 
             $D_{0123}$, $D_{0124}$, and $E_{01234}$.}
  \label{fig:1}
  \end{center}
\end{figure}
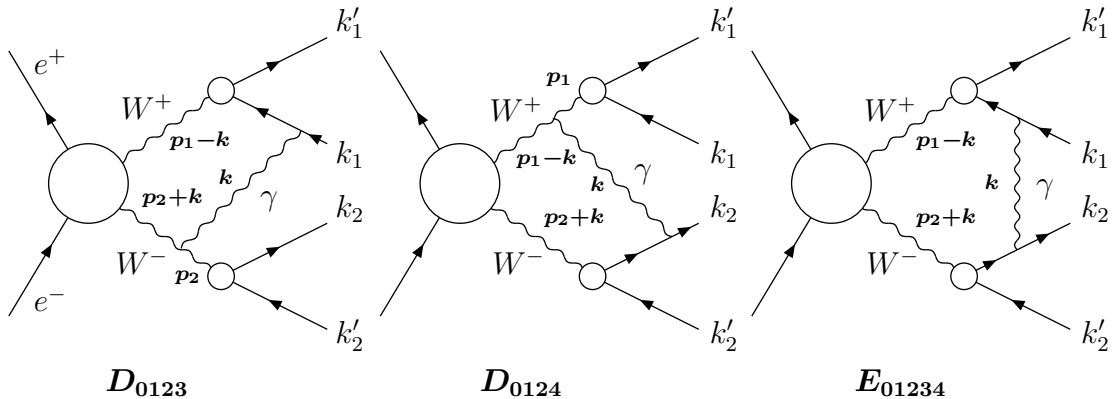
In order to give a gauge-invariant definition of the non-factorizable QED
corrections in the soft limit, we first restrict ourselves to the manifestly 
non-factorizable corrections, i.e.~those not having two resonant 
$W$-propagators as 
explicit overall factors. In addition we will restrict ourselves to the 
simplest 
class of charged-current 
four-fermion processes, involving a purely leptonic final state:
\begin{equation}
  e^+(q_1) e^-(q_2) \to W^+(p_1) + W^-(p_2) 
        \to \nu_{\ell}(k_1^{\prime}) \ell^+(k_1)
          + \ell^{\prime\,-}(k_2)\bar{\nu}_{\ell^{\prime}}(k_2^{\prime}).
\end{equation}
Whenever possible, all external fermions are taken to be massless. The relevant
contributions consist of the final--final and intermediate--final state 
photonic interactions displayed in Fig.~\ref{fig:1}. In principle also 
manifestly non-factorizable vertex corrections exist, which arise when the 
photon in Fig.~\ref{fig:1} does not originate from a $W$-boson line but from 
the $\gamma WW/ZWW$ vertex (hidden in the central blob). Those contributions 
can be shown to vanish in the double-pole approximation, using power-counting 
arguments \cite{review}. Also the manifestly non-factorizable initial--final 
state interference effects disappear in our approach. This happens upon adding 
virtual and real corrections, as will be briefly explained later. 

The double-pole contribution of the virtual corrections to the differential 
cross-section can be written in the form
\begin{equation}
\label{eq:mtrx_element}
  d\sigma_{\mbox{\scriptsize virt}} = 32\pi \alpha\, \real \biggl[
                         i (p_2 \cdot k_1) D_1 D_{0123}
                       + i (p_1 \cdot k_2) D_2 D_{0124} 
                       + i (k_1 \cdot k_2) D_1 D_2 E_{01234}
                       \biggr]\,d\sigma_{\mbox{\scriptsize Born}},
\end{equation}
where $D_{1,2} = p_{1,2}^2 - M_W^2 + i M_W\Gamma_W$ are the inverse 
(Breit--Wigner) $W$-boson propagators. The functions $D_{0123}$, $D_{0124}$, 
and $E_{01234}$ are the scalar integrals corresponding to the diagrams shown in
Fig.~\ref{fig:1}, with the integration measure defined as $d^4k/(2\pi)^4$.
The propagators occurring in these integrals are labelled according to:
0 = photon, 1 = $W^+$, 2 = $W^-$, 3 =  $\ell^+$, and 4 = $\ell^-$.
Note that the factorization property exhibited in Eq.\,(\ref{eq:mtrx_element})
is a direct consequence of the soft-photon approximation, which is inherent in
our approach. As a result, the propagators hidden inside the central blobs of
Fig.~\ref{fig:1} are Born-like, i.e.~unaffected by the presence of the 
non-factorizable photonic interactions.

In a similar way, only interferences of the real-photon diagrams can give 
contributions to the manifestly non-factorizable corrections. The relevant 
interferences can be read off from
Fig.~\ref{fig:1} by taking the exchanged photon to be on-shell. The infrared 
divergences contained in the virtual corrections will cancel against those 
present in the corresponding bremsstrahlung interferences.

\begin{figure}
  \unitlength 0.95pt
  \begin{center}
  \begin{picture}(180,150)(0,0)
    \ArrowLine(30,75)(0,125)
    \ArrowLine(0,25)(30,75)
    \Photon(30,75)(120,110){1}{9} 
    \Photon(30,75)(120,40){1}{9}  
    \ArrowLine(160,90)(120,110) 
    \Photon(105,45.8)(105,104.2){1}{5}           
    \ArrowLine(120,110)(160,130)      
    \ArrowLine(160,20)(120,40)      
    \ArrowLine(120,40)(160,60)      
    \GCirc(30,75){15}{1}
    \GCirc(120,110){5}{1} 
    \GCirc(120,40){5}{1}  
    \Text(75,0)[]{\boldmath $C_{012}$}
    \Text(10,123)[lb]{$e^+$}
    \Text(10,38)[lt]{$e^-$}
    \Text(70,105)[b]{$W^+$}
    \Text(70,53)[t]{$W^-$}
    \Text(117,85)[lt]{$\gamma$}
    \Text(172,137)[lb]{$k_1^{\prime}$}
    \Text(172,96)[lt]{$k_1$}
    \Text(172,64)[lb]{$k_2$}
    \Text(172,25)[lt]{$k_2^{\prime}$}
    \Text(106,79)[rb]{\boldmath $\scriptstyle k$}
    \Text(70,68)[lb]{\boldmath $\scriptstyle p_2+k$}
    \Text(70,94)[lt]{\boldmath $\scriptstyle p_1-k$}
    \Text(108,34)[lb]{\boldmath $\scriptstyle p_2$}
    \Text(108,119)[lb]{\boldmath $\scriptstyle p_1$}
  \end{picture} 
  \caption[]{The gauge-restoring ``Coulomb'' contribution. The corresponding
             scalar function is denoted by $C_{012}$. In 
             Sect.\,\ref{sec:feynman/coulomb} we shall briefly indicate the 
distinction
             between our ``Coulomb'' contribution, valid outside the 
             threshold region, and the usual one, which is also valid inside 
that region.}
  \label{fig:2}
  \end{center}
\end{figure}
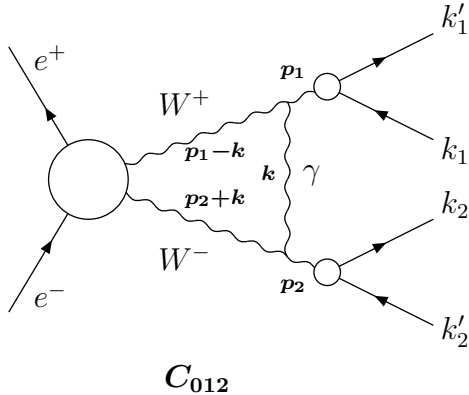

The set of manifestly non-factorizable QED diagrams displayed in 
Fig.~\ref{fig:1} is not $U(1)$ gauge invariant, which can be explicitly seen 
from
the non-vanishing of gauge-parameter-dependent terms when 
a general covariant gauge is used for the photon propagators. 
In order to achieve a gauge-invariant
definition of the non-factorizable corrections, all (soft) photonic 
interactions between the positively ($e^+,W^+,\ell^+$) and negatively
($e^-,W^-,\ell^{\prime\, -}$) charged particles should be taken into account,
which also holds for hadronic final states.
Looking at Fig.~\ref{fig:1}, this is equivalent to the set of all 
up--down QED interferences. 
The gauge invariance of these interference effects is a direct consequence
of the fact that in the soft-photon approximation the matrix element can be 
viewed as a product of two conserved currents: one, where the soft photon is 
attached
to the positively charged particles, and one, where it is attached to the 
negatively charged ones.
Only three of those up--down QED interferences are already present 
in Fig.\,\ref{fig:1}, 
all others except one will vanish in the soft-photon approximation.
In the soft-photon, double-pole approximation it is 
the ``Coulomb'' interaction between the off-shell $W$ bosons that survives 
as an 
extra contribution to the differential cross-section (see Fig.~\ref{fig:2}):
\begin{equation}
\label{Coulvirt}
  d\sigma^{\mbox{\scriptsize C}}_{\mbox{\scriptsize virt}}(p_1|p_2) =
  32\pi \alpha\, \real \biggl[ i (p_1 \cdot p_2) C_{012} \biggr]\,
  d\sigma_{\mbox{\scriptsize Born}}.
\end{equation}
The scalar three-point function $C_{012}$ is defined according to the 
above-defined notation. The terminology ``Coulomb'' interaction should not 
lead 
to confusion. It is a contribution that is a part of the diagram in 
Fig.\,\ref{fig:2}.
In Sect.\,\ref{sec:feynman/coulomb} we shall briefly indicate the distinction
between our ``Coulomb'' contribution, valid outside the threshold region,
and the usual one, which is primarily valid inside that region.

{}From the diagrams in Fig.~\ref{fig:1} it is clear that we have to 
calculate four- and five-point scalar functions and related
bremsstrahlung interference expressions, all in the soft-photon approximation.
In the next two sections we focus on the analytical results as obtained with 
the modified standard technique (MST), consisting of various elements. 
In particular the decomposition of the virtual and real five-point functions 
into a sum of four-point functions is given some special attention in 
Sect.\,\ref{sec:decomp}.\footnote{For higher $n$-point functions ($n>5$) this 
    decomposition can be carried out in an analogous way. Thus in principle 
    the methods outlined in Sects.\,\ref{sec:decomp/virt} and 
    \ref{sec:decomp/real} provide the basic tools for considering more 
    involved non-factorizable corrections, e.g.~for six-fermion final states.}
As such, the basic building blocks of the MST are the 
four-point functions $D$ and the related bremsstrahlung interference terms 
$D^{\mbox{\scriptsize R}}$. A general relation between the two entities is 
discussed in Sect.\,\ref{sec:feynman/prescriptions}. 
As a final step we derive in Sect.\,\ref{sec:feynman} the relevant scalar 
four-point functions $D$ in the soft-photon approximation by applying the 
Feynman-parameter technique. The related $D^{\mbox{\scriptsize R}}$ functions 
are obtained by using a ``particle-pole'' expression and performing certain 
substitutions.
 
In Sect.\,\ref{sec:dmi} we present the calculation along the lines of 
Ref.\,\cite{melyak}, which involves the method of direct momentum integration. 
This calculation will serve as a check of the MST results. Moreover, it is
required for pinpointing the source of the differences with \cite{melyak}.
Whereas the method of Sects.\,\ref{sec:decomp} and \ref{sec:feynman} seems to 
be quite general, the method of Sect.\,\ref{sec:dmi} is unlikely to be 
applied to $n$-point functions with $n>5$. As we will see, it just becomes 
too complicated.

In Sect.\,\ref{sec:results} we present the main analytical features of our 
study and we indicate how the results for semileptonic and hadronic final 
states can be obtained from the purely leptonic case. For the numerical 
implications of the non-factorizable corrections we
refer to \cite{nfletter}. Our calculations confirm that non-factorizable 
corrections vanish in the special case of initial--final state interference,
thereby making non-factorizable radiative corrections independent from the 
$W$ production angle, and in all cases when the integrations over both 
invariant masses of the virtual $W$ bosons are performed \cite{fadin-khoze}. 
The practical
consequence of the latter is that pure angular distributions are unaffected by
non-factorizable $\cal{O}(\alpha)$ corrections. So, the studies of non-Abelian
triple gauge-boson couplings at LEP2 \cite{anom-coupling} are not affected
by these corrections. The non-factorizable $\cal{O}(\alpha)$ corrections,
however, do affect the invariant-mass distributions ($W$ line-shapes).
These distributions play a crucial role in 
extracting the $W$-boson mass from the data through direct reconstruction of 
the Breit--Wigner resonances. The non-factorizable corrections to the 
line-shapes turn out to be the same for quark and lepton final states, 
provided the integrations over the decay angles have been performed.
Finally, in Sect.\,\ref{conclusions} we draw some conclusions.

\section{Modified standard technique: basic ingredients}
\label{sec:decomp}

In this section the basic ingredients are presented for the evaluation of 
non-factorizable corrections in the MST. As a 
first step we discuss the decomposition of virtual and real five-point 
functions into a sum of four-point functions. Subsequently we demonstrate how
virtual and real contributions can be related in the soft-photon approximation.
Having established the five-point decompositions and the relation between 
virtual and real contributions, the actual calculation in the MST boils down
to the evaluation of scalar four-point integrals.

\subsection{Decomposition of the virtual five-point function}
\label{sec:decomp/virt}

In this subsection we derive the decomposition of the virtual scalar five-point
function into a sum of scalar four-point functions. The derivation follows 
Ref.\,\cite{neerven}. The reason for repeating this calculation lies in the 
fact that it will serve as guideline for the decomposition of the real 
five-point function, which has not been considered before.

Let us consider the following general five-point function:
\be
  E_{01234} = \int \frac{d^{4}k}{(2\pi)^{4}}\,
              \frac{1}{N_{0} N_{1} N_{2} N_{3} N_{4}},
\ee
where
\be
  N_{0} = k^{2} - \lambda^{2} + io\ \ \ \mbox{and} \ \ \ 
  N_{i} = (k + p_{i})^{2} - m_{i}^{2} + io.
\ee
Here $io$ denotes an infinitesimal imaginary part. The plus sign accompanying
this imaginary part is determined by causality. The mass parameter $\lambda$
is in principle arbitrary. In our case, however, it will denote a non-zero
photon mass, needed for regularizing the infrared divergences.

Before starting with the decomposition, we first derive a useful identity.
To this end we exploit Lorentz covariance and write
\be
\label{lorentzcov}
  \int \frac{d^{4}k}{(2\pi)^{4}}\,\frac{k^{\mu}}{N_{0} N_{1} N_{2} N_{3}}
               =  c_1\,p_{1}^{\mu} + c_2\,p_{2}^{\mu} + c_3\,p_{3}^{\mu}.
\ee
The integral on the left-hand side is ultraviolet-finite and, when properly 
regularized, also infrared-finite. The quantities $c_i$ on the right-hand 
side are therefore finite coefficients, dependent on masses and the invariants
$p_i\cdot p_j$ ($i,j=1,2,3$). Contracting this expression with the
antisymmetric Levi--Civita tensor, one obtains the identity
\be
\label{eq:decomp/theorem-virt}
  \int \frac{d^{4}k}{(2\pi)^{4}}\,
       \frac{\varepsilon_{p_{1}p_{2}p_{3}k}}{N_{0} N_{1} N_{2} N_{3}} = 0,
\ee
where we introduced the widely-used notation 
\be
  \varepsilon_{\mu\nu\rho p} = \varepsilon_{\mu\nu\rho\sigma} p^{\sigma}, 
  \ \ \ 
  \varepsilon_{\mu\nu p q} = \varepsilon_{\mu\nu\rho\sigma} p^{\rho}q^{\sigma},
  \ \cdots
\ee
In our convention, the Levi--Civita tensor is defined according to 
$\varepsilon^{0123} = -\varepsilon_{0123} = 1$. 
The above identity will prove extremely useful in the following.
 
The actual derivation of the decomposition formula starts with the Schouten 
identity
\be
  a\,k^{\mu} = \sum_{i=1}^4 v_{i}^{\mu}\,(p_{i}\cdot k), 
\ee
where the following notation was used:
\be
  v_{1}^{\mu} = \varepsilon^{\mu p_{2} p_{3} p_{4}}, \ \ \ 
  v_{2}^{\mu} = \varepsilon^{p_{1} \mu p_{3} p_{4}}, \ \ \  
  v_{3}^{\mu} = \varepsilon^{p_{1} p_{2} \mu p_{4}}, \ \ \ 
  v_{4}^{\mu} = \varepsilon^{p_{1} p_{2} p_{3} \mu}, \ \ \ 
  a = \varepsilon_{p_{1}p_{2}p_{3}p_{4}} = \varepsilon^{p_{1}p_{2}p_{3}p_{4}}.
\ee
Note that from the quantity $a$ one can construct the Gram-determinant of the 
system, $\Delta_{4} = a^2$. The next step in the derivation is to contract the 
Schouten identity with $k_{\mu}$, yielding 
\be
\label{eq:decomp/virt/schouten}
  a\,k^{2} = \sum_{i=1}^4 (k\cdot v_{i})(k\cdot p_{i}).
\ee
Now we can substitute 
\begin{eqnarray}
\label{substitution}
  k^{2} &=& N_{0} + \lambda^{2} - io, \nonumber \\[1mm]
  (k\cdot p_{i}) &=& \frac{1}{2}\, [N_{i} - N_{0} - r_{i}], \ \ \mbox{with}\ \ 
                     r_{i} = p_{i}^{2} - m_{i}^{2} + \lambda^{2}
\end{eqnarray}
to arrive at
\be
  2 a\,(N_{0}+\lambda^{2}) = \sum_{i=1}^4 (k\cdot v_{i}) (N_{i}-N_{0}-r_{i}).
\ee
In order to make the link to the scalar five-point function, one should divide
this expression by $N_{0} N_{1} N_{2} N_{3} N_{4}$ and perform the integration 
over $d^{4}k$. As a result of Eq.\,(\ref{eq:decomp/theorem-virt}) the $N_{i}$ 
terms vanish. The terms $\sum (k\cdot v_{i})\,N_{0}$ can be transformed
according to 
\be
\label{kvi}
  \sum_{i=1}^4 (k\cdot v_{i}) = 
      \varepsilon^{(k+p_{1})(p_{2}-p_{1})(p_{3}-p_{1})(p_{4}-p_{1})} - a,
\ee
which can be verified by a direct check. After integration the first term will
vanish. This can be most easily seen by making a change of integration
variable, $\tilde{k}^{\mu}=k^{\mu}+p_{1}^{\mu}$, and subsequently applying
Eq.\,(\ref{eq:decomp/theorem-virt}).
The complete expression now reads
\be
  \int \frac{d^{4}k}{(2\pi)^{4}}\,
       \frac{a N_{0} + 2 a \lambda^{2} + \sum r_{i} (k\cdot v_{i})}
            {N_{0} N_{1} N_{2} N_{3} N_{4}} =0.
\ee

The final step is to multiply this expression by $a$ and to apply the Schouten
identity and Eq.\,(\ref{substitution}) to the last term in the numerator.
This allows us to express the complete numerator in terms of the propagators
appearing in the denominator:
\be
  \int \frac{d^{4}k}{(2\pi)^{4}}\,
       \frac{2 \lambda^{2}\Delta_{4} - \frac{1}{2}w^{2} + N_{0} \Delta_{4}
             - \frac{1}{2} N_{0}\sum(v_{i}\cdot w) 
             + \frac{1}{2}\sum N_{i}(v_{i}\cdot w)}
            {N_{0} N_{1} N_{2} N_{3} N_{4}} = 0,
\ee
with
\be
  w^{\mu} = \sum_{j=1}^4 r_{j}\,v_{j}^{\mu}.
\ee
The final formula for the decomposition reads
\begin{eqnarray}
\label{eq:decomp/virt}
  (w^{2} - 4 \lambda^{2} \Delta_{4})\,E_{01234} &=&
        (w \cdot v_1)\,D_{0234} + (w \cdot v_2)\,D_{0134} 
      + (w \cdot v_3)\,D_{0124} + (w \cdot v_4)\,D_{0123} \nonumber \\[1mm]
                                                & &
      {}+ \biggl[ 2\Delta_{4} - \sum_{i=1}^4 (w\cdot v_{i}) \biggr]\,D_{1234},
\end{eqnarray}
where $D_{0234},\,D_{0134}$, etc.,~denote four-point scalar functions 
containing
the propagators with labels $(0,2,3,4)$, $(0,1,3,4)$, etc. 

The generalization of this decomposition to higher multipoint functions can be 
performed in a similar way \cite{neerven}. In general, a scalar $N$-point 
function can be expressed in terms of the $N$ underlying $(N-1)$-point 
functions.

\subsection{Decomposition of the real five-point function}
\label{sec:decomp/real}

Using the derivation presented in the previous subsection as guideline, we can 
now try to derive a similar decomposition for the real five-point function. 
As can be read off from Fig.~\ref{fig:1}, by taking the exchanged photon to be 
on-shell, the real five-point function takes the form
\be
  E_{01234}^{\mbox{\scriptsize R}} = \int \frac{d^{3}k}{(2\pi)^{3}\,2\omega}\, 
           \frac{1}{N_{1}^{\prime}N_{2}^{\prime}N_{3}^{\prime}N_{4}^{\prime}},
\ee
where 
\be
  \omega=\sqrt{\vec{k}^{2}+\lambda^{2}}\, , \ \ \ 
  N_{1,2}^{\prime} = N_{1,2}, \ \ \mbox{and} \ \ 
  N_{3,4}^{\prime} = N_{3,4}^{*}.
\ee
The photon is now on-shell, so $k^{2} = \lambda^{2}$ and $N_{0}=0$. Note that 
the momenta $p_i$, hidden inside $N_i$, are time-like and have positive energy 
components.

One can proceed in the same way as in the case of the decomposition of the
virtual five-point function. The Schouten identity is still valid, but 
Eq.\,(\ref{eq:decomp/theorem-virt}) in its old form does not work in the case 
of real-photon radiation, and should be modified. In the derivation of the 
virtual decomposition, Eq.\,(\ref{eq:decomp/theorem-virt}) was used twice, 
leading to the nullification of 
\be
  \int\frac{d^{4}k}{(2\pi)^{4}}\, 
          \frac{\sum (k\cdot v_{i}) N_{i}}{N_0 N_{1} N_{2} N_{3} N_{4}} 
          \ \ \ \mbox{and} \ \ \ 
  \int\frac{d^{4}k}{(2\pi)^{4}}\,
          \frac{N_{0}\,[\sum(k\cdot v_{i}) + a]}{N_0 N_{1} N_{2} N_{3} N_{4}}.
\ee
In the case of real-photon radiation, this will correspond to 
\be
  \int\frac{d^{3}k}{(2\pi)^{3}\,2\omega}\,
          \frac{\sum (k\cdot v_{i})\,N_{i}^{\prime}}
               {N_{1}^{\prime}N_{2}^{\prime}N_{3}^{\prime}N_{4}^{\prime}}
          \ \ \ \mbox{and} \ \ \ 
  \int\frac{d^{3}k}{(2\pi)^{3}\,2\omega}\,
          \frac{N_{0}\,[\sum(k\cdot v_{i}) + a]}
               {N_{1}^{\prime}N_{2}^{\prime}N_{3}^{\prime}N_{4}^{\prime}}.
\ee
For the nullification of the second integral the validity of
Eq.\,(\ref{eq:decomp/theorem-virt}) is immaterial, since for the on-shell 
photon $N_{0}=0$ anyway. The first integral, however, is no longer 
necessarily zero.
The fact that the photon is on-shell implies that $k^{2}=\lambda^{2}$
and that the propagators $N_{i}$ are linear in $k$. By simple power counting, 
one can conclude that this integral is formally ultraviolet-divergent. 
For this reason, the Lorentz-covariance argument used in 
Eq.\,(\ref{lorentzcov}) is not correct any more and 
Eq.\,(\ref{eq:decomp/theorem-virt}) is invalidated.

Apart from the modification of Eq.\,(\ref{eq:decomp/theorem-virt}) and the 
fact that $N_{0}=0$, the derivation of the decomposition for the real 
five-point function is not changed, resulting in
\begin{eqnarray}
\label{eq:decomp/real/1}
  (w^{2} - 4\lambda^{2}\Delta_{4})\,E_{01234}^{\mbox{\scriptsize R}} &=&
            (w \cdot v_1)\,D_{0234}^{\mbox{\scriptsize R}} 
          + (w \cdot v_2)\,D_{0134}^{\mbox{\scriptsize R}} 
          + (w \cdot v_3)\,D_{0124}^{\mbox{\scriptsize R}}
          + (w \cdot v_4)\,D_{0123}^{\mbox{\scriptsize R}} \nonumber \\[1mm]
                                                                     & &
          {}- 2\int\frac{d^{3}k}{(2\pi)^{3}\,2\omega}\,
                 \frac{\sum a\,(k\cdot v_{i})\,N_{i}^{\prime}}
                 {N_{1}^{\prime}N_{2}^{\prime}N_{3}^{\prime}N_{4}^{\prime}}.
\end{eqnarray}

The main difference with the virtual decomposition is the occurrence of the 
last term in Eq.\,(\ref{eq:decomp/real/1}). It turns out that the poles in 
this 
particular integral can be moved in such a way that 
$N_{i}^{\prime}\to N_{i}$ for all $i$. Indeed, the integral can be rewritten
in the following way:
\be
  \int \frac{d^{3}k}{(2\pi)^{3}\,2\omega}\,
       \frac{\sum (k\cdot v_{i})\,N_{i}^{\prime}}
            {N_{1}^{\prime}N_{2}^{\prime}N_{3}^{\prime}N_{4}^{\prime}}
  =
  \int \frac{d^{3}k}{(2\pi)^{3}\,2\omega}\,
       \frac{\sum (k\cdot v_{i})\,N_{i}}{N_{1}N_{2}N_{3}N_{4}}
        + \Delta^{(1)} + \Delta^{(2)},
\ee
with
\begin{eqnarray}
  \Delta^{(1)} &=& \int \frac{d^{3}k}{(2\pi)^{3}\,2\omega}\,
            \frac{(k\cdot v_{1})\,N_{1}+(k\cdot v_{2})\,N_{2}}{N_{1}N_{2}}
            \Bigl[\frac{1}{N_{3}^{\prime}N_{4}^{\prime}}
                 -\frac{1}{N_{3}N_{4}}
            \Bigr], \nonumber \\[1mm]
  \Delta^{(2)} &=& \int \frac{d^{3}k}{(2\pi)^{3}\,2\omega}\,
            \Biggl\{ \frac{(k\cdot v_{3})}{N_{1}N_{2}}
                     \Bigl[ \frac{1}{N_{4}^{\prime}} - \frac{1}{N_{4}} \Bigr]
                   + \frac{(k\cdot v_{4})}{N_{1}N_{2}}
                     \Bigl[ \frac{1}{N_{3}^{\prime}} - \frac{1}{N_{3}} \Bigr] 
            \Biggr\}.
\end{eqnarray}
Both $\Delta^{(1)}$ and $\Delta^{(2)}$ are in fact zero. Let us consider, for
example, one of the terms contributing to 
$\Delta^{(2)}=\Delta^{(2)}_{3}+\Delta^{(2)}_{4}$, e.g.
\be
  \Delta^{(2)}_{4} = \int\frac{d^{3}k}{(2\pi)^{3}\,2\omega}\,
                         \frac{(k\cdot v_{4})}{N_{1}N_{2}}\,
                         \frac{2 i\, \imag N_3}{N_{3}N_{3}^{\prime}}.
\ee
This integral is ultraviolet-finite, even for an on-shell photon, and therefore
no regularization is needed. Consequently, the Lorentz-covariance argument is 
valid:
\be
  \int\frac{d^{3}k}{2\omega}\,\frac{k^{\mu}}{N_{1}N_{2}N_{3}N_{3}^{\prime}}
  = c_1\,p_{1}^{\mu} + c_2\,p_{2}^{\mu} + c_3\,p_{3}^{\mu},
\ee
where the quantities $c_i$ are finite coefficients. Contracting the last 
expression with $v_{4\,\mu}=\varepsilon_{p_{1}p_{2}p_{3}\mu}$, one arrives at
$\Delta^{(2)}_4 = 0$. Using similar arguments one can prove that 
$\Delta^{(1)}=\Delta^{(2)}=0$. 

An important point in this line of reasoning was the use of Lorentz-invariance 
of the integration $d^{3}k/\omega$. Such an integration is indeed 
Lorentz-invariant, provided that the area of integration is Lorentz-invariant. 
In the context of the double-pole approximation, the photon is treated 
inclusively, with the integration performed over all possible values of $k$ up 
to infinity. If one would, however, consider an exclusive process, involving
the introduction of a cutoff $\Omega_{max}$, then the area of integration 
might fail to be Lorentz-invariant, and the decomposition stops at 
Eq.(\ref{eq:decomp/real/1}). In order to successfully proceed beyond that point
for exclusive processes, one should make sure that the cut-off prescription, 
which defines the area of integration, does not introduce new independent 
four-vectors in the integral. If this condition is satisfied, a generalization
of the decomposition to exclusive bremsstrahlung processes should be feasible.

So, in our approach, the following identity has been established:
\be
\label{eq:decomp/real/move_the_poles}
  \int \frac{d^{3}k}{(2\pi)^{3}\,2\omega}\,
       \frac{\sum (k\cdot v_{i})\,N_{i}^{\prime}}
            {N_{1}^{\prime}N_{2}^{\prime}N_{3}^{\prime}N_{4}^{\prime}}
  =
  \int \frac{d^{3}k}{(2\pi)^{3}\,2\omega}\,
       \frac{\sum (k\cdot v_{i})\,N_{i}}{N_{1}N_{2}N_{3}N_{4}}.
\ee
As was already noted, the integral on the right-hand side is formally 
divergent. Its virtual analogue, being formally finite, vanishes:
\be
  \int \frac{d^{4}k}{(2\pi)^{4}}\,
       \frac{\sum (k\cdot v_{i})\,N_{i}}{N_{0}N_{1}N_{2}N_{3}N_{4}} = 0.
\ee
Performing a contour integration in the lower half of the complex $k_0$-plane, 
indicated by the integration contour $C_O$ in Fig.~\ref{fig:decomp1}, one
can use this identity to relate the photon-pole contribution to the 
particle-pole contributions: 
\begin{figure}[t]
  \unitlength 1cm
  \begin{center}
  \begin{picture}(17,8)
  \put(12,6){\makebox[0pt][c]{$(k_0)$}}
  \put(9.5,0.5){\makebox[0pt][c]{$C_{O}$}}
  \put(4,-3){\includegraphics{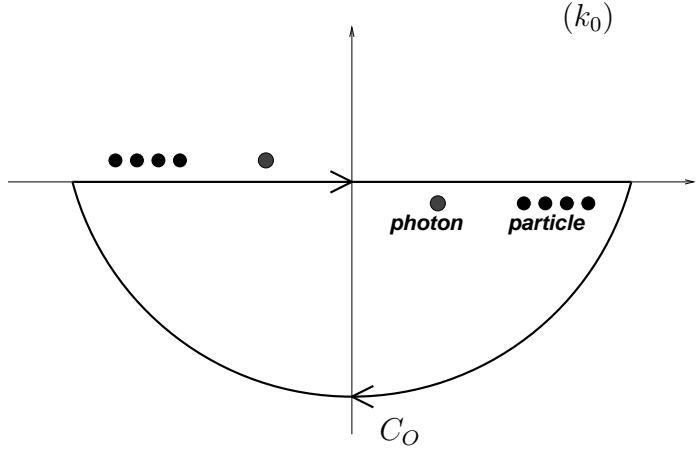}}
  \end{picture}
  \end{center}
  \caption[]{Integration contour in the lower half of the complex $k_0$-plane 
leading 
             to Eq.\,(\protect\ref{contourint}).}
\label{fig:decomp1}
\end{figure}
\be
\label{contourint}
  -2\pi i \int\frac{d^{3}k}{(2\pi)^{4}\,2\omega}\,
               \frac{\sum (k\cdot v_{i}) N_{i}}{N_{1}N_{2}N_{3}N_{4}}
  =
  - \int\frac{d^{4}k}{(2\pi)^{4}N_{0}}\,\pole\,
        \frac{\sum (k\cdot v_{i}) N_{i}}{N_{1}N_{2}N_{3}N_{4}}.
\ee
Here ``$\pole$'' denotes the complex particle poles that should be taken into 
account. Note that the left-hand side of Eq.\,(\ref{contourint}) corresponds
to the real-photon radiation integral that we are pursuing to evaluate. The
integral on the right-hand side does not correspond to on-shell photons any 
more, since $N_0 \neq 0$.
Now we can use the Schouten identity to substitute
\be
  \sum_{i=1}^4 (k\cdot v_{i})\,N_{i} = \sum_{i=1}^4 (k\cdot v_{i})\,N_{0} 
              + 2 a k^{2} + \sum_{i=1}^4 r_{i}\,(k\cdot v_{i})
\ee
on the right-hand side. After some rearrangements one obtains
\be
  2\pi i \int\frac{d^{3}k}{(2\pi)^{4}\,2\omega}\,
             \frac{\sum (k\cdot v_{i})\,N_{i}}{N_{1}N_{2}N_{3}N_{4}}
  =
    \int\frac{d^{4}k}{(2\pi)^{4}}\,
        \frac{\sum (k\cdot v_{i})+2a}{N_{1}N_{2}N_{3}N_{4}}
  + \int\frac{d^{4}k}{(2\pi)^{4}N_{0}}\,\pole\,
        \frac{\sum r_{i}\,(k\cdot v_{i}) + 2 a \lambda^{2}}
             {N_{1}N_{2}N_{3}N_{4}}.
\ee
The first integral on the right-hand side can be simplified with the help of
Eqs.\,(\ref{kvi}) and (\ref{eq:decomp/theorem-virt}), yielding
\be
\label{eq:decomp/real/2}
  2\pi i \int\frac{d^{3}k}{(2\pi)^{4}\,2\omega}\,
             \frac{\sum (k\cdot v_{i})\,N_{i}}{N_{1}N_{2}N_{3}N_{4}}
  =
  a \int\frac{d^{4}k}{(2\pi)^{4}}\,\frac{1}{N_{1}N_{2}N_{3}N_{4}} + {\cal R},
\ee
with
\be
  {\cal R} = \int\frac{d^{4}k}{(2\pi)^{4}N_{0}}\,\pole\,
                 \frac{\sum r_{i}\,(k\cdot v_{i}) + 2 a \lambda^{2}}
                      {N_{1}N_{2}N_{3}N_{4}}.
\ee
In App.\,\ref{app:decomp} it is shown that the ${\cal R}$ term in 
(\ref{eq:decomp/real/2}), which consists of a combination of particle-pole
contributions, vanishes. After that the decomposition for the real five-point 
function can be written in a compact form, analogous to the decomposition 
of the virtual five-point function
\begin{eqnarray}
\label{eq:decomp/real}
  (w^{2} - 4\lambda^{2}\Delta_{4})\,E_{01234}^{\mbox{\scriptsize R}} &=&
            (w \cdot v_1)\,D_{0234}^{\mbox{\scriptsize R}} 
          + (w \cdot v_2)\,D_{0134}^{\mbox{\scriptsize R}} 
          + (w \cdot v_3)\,D_{0124}^{\mbox{\scriptsize R}}
          + (w \cdot v_4)\,D_{0123}^{\mbox{\scriptsize R}} \nonumber \\[1mm]
                                                                     & &
          {}+ 2 i \Delta_{4}\int \frac{d^{4}k}{(2\pi)^{4}}\,
                               \frac{1}{N_{1}N_{2}N_{3}N_{4}}.
\end{eqnarray}
Note that the last term on the right-hand side of Eq.\,(\ref{eq:decomp/real})
is exactly a virtual scalar four-point function with a coefficient 
$2 i\Delta_{4}$.
In comparing Eq.\,(\ref{eq:decomp/virt}) to Eq.\,(\ref{eq:decomp/real}) one 
observes
certain similarities: the first four terms in 
Eq.\,(\ref{eq:decomp/real}) are the radiative analogues of their virtual 
counterparts
in Eq.\,(\ref{eq:decomp/virt}). One may naively think that the last term in
(\ref{eq:decomp/real}) should not be there, since it does not correspond to 
photon radiation.
In fact, it is a direct consequence of having ultraviolet-divergent integrals
during the intermediate steps of the derivation.

This concludes the derivation of the decomposition of the five-point function
corresponding to inclusive bremsstrahlung. As was noted before, generalization
to the case of exclusive bremsstrahlung is possible, provided that the cut-off
is introduced in such a way that no new independent four-vectors appear in 
the integrals. In analogy to what was remarked for the virtual decomposition,
also the generalization to higher multipoint radiation functions is possible
and rather straightforward. One should simply follow the approach of 
Ref.\,\cite{neerven} for multipoint scalar functions.

\subsection{Application of the five-point decompositions}
\label{sec:decomp/application}

We can now apply the five-point decompositions to the non-factorizable 
$W$-pair corrections. The virtual scalar five-point function, corresponding 
to the third diagram in Fig.~\ref{fig:1}, reads in the double-pole 
approximation
\be
\label{eq:decomp/E01234}
  w^2\,E_{01234} = 2 \Delta_{4}\,D_{1234}   
                 + (w \cdot v_1)\,D_{0234} + (w \cdot v_2)\,D_{0134} 
                 + (w \cdot v_3)\,D_{0124} + (w \cdot v_4)\,D_{0123},
\ee
with 
\begin{eqnarray} 
          v_{1 \mu} &=& -\,\varepsilon_{\mu p_2 k_1 k_2}, \ \ \ \ \ \ \ \ 
          v_{2 \mu}  =  +\,\varepsilon_{p_1 \mu k_1 k_2}, \ \ \ \ \ \ \ \ 
          v_{3 \mu}  =  -\,\varepsilon_{p_1 p_2 \mu k_2}, \nonumber \\
          v_{4 \mu} &=& +\,\varepsilon_{p_1 p_2 k_1 \mu}, \ \ \ \ \ \ \ \ 
          w^{\mu}    =  D_1 v_1^{\mu} + D_2 v_2^{\mu}, \ \ \ \;
          \Delta_4 = [\, \varepsilon_{p_1 p_2 k_1 k_2}\, ]^2.
\end{eqnarray}
Comparison with Eq.\,(\ref{eq:decomp/virt}) reveals that the terms 
$-\sum (w\cdot v_{i})\,D_{1234}$ have been neglected, since they are formally
of higher order in the expansion in powers of $\Gamma_W/M_W$. Note that the 
scalar four-point function $D_{1234}$ is purely a consequence of the 
decomposition (\ref{eq:decomp/E01234}). It does not involve the exchange of a 
photon and is therefore not affected by the soft-photon approximation. 
Since the factor $2 \Delta_{4}/w^2$ is already doubly resonant, $D_{1234}$ 
should be calculated for on-shell $W$ bosons. The scalar four-point functions  
$D_{0134}$ and $D_{0234}$ are infrared-divergent and should be calculated in 
the soft-photon approximation.

In the same way, the five-point function corresponding to real-photon radiation
is given by
\be
\label{eq:decomp/E01234real}
  w^{\prime 2}\,E^{\mbox{\scriptsize R}}_{01234} = 
       2i \Delta_{4}\,D^{\mbox{\scriptsize R}}_{1234}
     + (w^{\prime} \cdot v_1^{\prime})\,D^{\mbox{\scriptsize R}}_{0234} 
     + (w^{\prime} \cdot v_2^{\prime})\,D^{\mbox{\scriptsize R}}_{0134} 
     + (w^{\prime} \cdot v_3^{\prime})\,D^{\mbox{\scriptsize R}}_{0124} 
     + (w^{\prime} \cdot v_4^{\prime})\,D^{\mbox{\scriptsize R}}_{0123}.
\ee
The four-vectors $w^{\prime}$ and $v_{i}^{\prime}$ are defined as before, but 
for real-photon emission. This is equivalent to the following substitutions: 
$p_1 \to -p_1$, $k_1 \to -k_1$ and $D_2 \to D_2^*$. The radiation function
$D_{1234}^{\mbox{\scriptsize R}}$ is an artefact of the decomposition 
(\ref{eq:decomp/E01234real}) and does not involve the exchange of a photon. 
It can be obtained from $D_{1234}$ by the substitutions $p_1 \to -p_1$ and 
$k_1 \to -k_1$. In the double-pole approximation, i.e.~for on-shell $W$ bosons,
this implies the relation  
$D_{1234}^{\mbox{\scriptsize R}} = i \imag D_{1234}$. This property ensures 
the 
cancellation of the virtual non-factorizable $D_{1234}$-dependent corrections
against the corresponding real-photon corrections, provided that the 
integration over the $W$-boson virtualities (i.e.~$D_{1,2}$) is performed. 
This phenomenon is a general consequence of the soft-photon approximation 
\cite{fadin-khoze}.

\subsection{Connection between virtual and real contributions}
\label{sec:feynman/prescriptions}

At this point we have reduced the calculation of the non-factorizable 
corrections to the evaluation of virtual and real four-point functions. We can,
however, go one step further and establish a connection between the 
contribution from the photon-pole part of the virtual scalar functions, 
$D^{\gamma}$, and the corresponding radiative interferences. To this end, we 
consider for example the contributions related to $D_{0123}^{\gamma}$ and 
$D_{0123}^{\mbox{\scriptsize R}}$. The contribution of the radiative 
interference to the cross-section (see Fig.~\ref{fig:1}) is given by
\be
\label{eq:feynman/prescriptions/1}
  d\sigma_{\mbox{\scriptsize real}}(D_{0123}^{\mbox{\scriptsize R}}) = 
      d\sigma_{\mbox{\scriptsize Born}}\,
      \real \!\int \frac{d^{3}k}{(2\pi)^{3}\,2\omega}\,
      \frac{32\pi\alpha\,(p_{2}\cdot k_{1})\,D_{1}}
           {[D_{1} + 2 (p_{1}\cdot k)] [D_{2}^{*} + 2 (p_{2}\cdot k)] 
            [2(k_{1}\cdot k) + io]},
\ee
where $k_{0}=\omega=|\vec{k}|$\,.

This has to be compared with the corresponding photon-pole part of the virtual 
correction. This contribution is evaluated in the lower half of the complex 
$k_0$-plane, where the photon pole is situated at $k_0=\omega=|\vec{k}|-io$:
\be
  d\sigma_{\mbox{\scriptsize virt}}(D_{0123}^{\gamma}) =
      d\sigma_{\mbox{\scriptsize Born}}\, 
      \real \!\int \frac{d^{3}k}{(2\pi)^{3}\,2\omega}\,
      \frac{32\pi\alpha\,(p_{2}\cdot k_{1})\,D_{1}}
           {[D_{1} - 2 (p_{1}\cdot k)] [D_{2} + 2 (p_{2}\cdot k)] 
            [ - 2(k_{1}\cdot k) + io]}.
\ee
This can be rewritten in the form
\be
\label{eq:feynman/prescriptions/2}
  d\sigma_{\mbox{\scriptsize virt}}(D_{0123}^{\gamma}) =
      d\sigma_{\mbox{\scriptsize Born}}\,
      \real \!\int \frac{d^{3}k}{(2\pi)^{3}\,2\omega}\,
      \frac{32\pi\alpha\,(p_{2}\cdot k_{1})\,D_{1}^{*}}
           {[D_{1}^{*} - 2 (p_{1}\cdot k)] [D_{2}^{*} + 2 (p_{2}\cdot k)] 
            [-2(k_{1}\cdot k) - io]}.
\end{equation}
Comparing Eqs.\,(\ref{eq:feynman/prescriptions/1}) and 
(\ref{eq:feynman/prescriptions/2}), one can readily see that 
$d\sigma_{\mbox{\scriptsize real}}(D_{0123}^{\mbox{\scriptsize R}})$ can be
obtained from the photon-pole contribution to the virtual correction  
$d\sigma_{\mbox{\scriptsize virt}}(D_{0123}^{\gamma})$, by adding an overall
minus sign and substituting\footnote{Note that 
   $d\sigma_{\mbox{\scriptsize Born}}$ is not affected by substitutions of the
   form $D_i \to -D_i^*$ \,$(i=1,2)$, since it only depends on $|D_1 D_2|^2$.} 
$D_{1}\to-D_{1}^{*}$. In a similar way 
$d\sigma_{\mbox{\scriptsize real}}(D_{0124}^{\mbox{\scriptsize R}})$ can be
obtained from the photon-pole contribution to the virtual correction  
$d\sigma_{\mbox{\scriptsize virt}}(D_{0124}^{\gamma})$, by adding an overall
minus sign and substituting $D_{2}\to-D_{2}^{*}$. The different substitution
rule reflects the fact that we will determine $D_{0124}$ and 
$D_{0124}^{\gamma}$ from $D_{0123}$ and $D_{0123}^{\gamma}$ by substituting
$(p_1,k_1) \leftrightarrow (p_2,k_2)$. Note that this is equivalent to
evaluating $D_{0124}^{\gamma}$ in the upper half of the complex $k_0$-plane.

Also the ``Coulomb'' and five-point contributions can be treated in this way,
bearing in mind that the coefficients of the five-point decomposition also
depend on $D_{1,2}$. In conclusion, the following relation emerges.
The radiative interferences can be obtained from Eqs.\,(\ref{eq:mtrx_element}) 
and (\ref{Coulvirt}) by adding a minus sign, by inserting the decomposition 
given in Eq.\,(\ref{eq:decomp/E01234}), and by substituting 
\begin{eqnarray*}
  \mbox{-- in the $D_{0123},D_{0134}$ terms:} &&  
       D_{0123},D_{0134} \to D^{\gamma}_{0123},D^{\gamma}_{0134}
       \mbox{ \ followed by \ } D_1 \to -D_1^*, \\
  \mbox{-- in the $D_{0124},D_{0234}$ terms:} &&  
       D_{0124},D_{0234} \to D^{\gamma}_{0124},D^{\gamma}_{0234}
       \mbox{ \ followed by \ } D_2 \to -D_2^*, \\
  \mbox{-- in the $D_{1234}$ terms:}\hphantom{,D_{0234}} &&  
       D_{1234} \to D^{\mbox{\scriptsize R}}_{1234}
       \hphantom{,D^{\gamma}_{0124},D^{\gamma}_{0234}} 
       \mbox{ \ followed by \ } D_2 \to -D_2^*, \\
  \mbox{-- in the $C_{012}$ terms: }\hphantom{,D_{0234}} &&  
       C_{012} \to C^{\gamma}_{012}
       \hphantom{,D^{\gamma}_{0124},D^{\gamma}_{0234}\ \ \ } 
       \mbox{ \ followed by \ } D_1 \to -D_1^*. 
\end{eqnarray*}
Here both $D_{0124}^{\gamma}$ and $D_{0234}^{\gamma}$ are determined by 
substituting $(p_1,k_1) \leftrightarrow (p_2,k_2)$ in the expressions for 
$D_{0123}^{\gamma}$ and $D_{0134}^{\gamma}$, respectively. As such, the above
connection between real and virtual corrections implies that 
$D_{0124}^{\gamma}$ and $D_{0234}^{\gamma}$ are evaluated in the upper 
half-plane.

Note that the above-presented connection between the virtual and real 
non-factorizable corrections hinges on two things. First of all, the inclusive 
treatment of the bremsstrahlung photon, with the phase-space integration 
extending to infinity. Second, the fact that both virtual and real 
corrections are 
calculated in the soft-photon approximation, inherent in the double-pole 
approach. 

As mentioned before, manifestly non-factorizable initial--final state 
interference effects are also possible in our approach. As stated in 
Sect.\,\ref{sec:gaugeinv} we will now
briefly indicate why these effects vanish. Let us consider, for example,
the initial--final state interference contribution corresponding to the
photonic interaction between the positron [$e^+(q_1)$] and the positively
charged final-state lepton [$\ell^+(k_1)$]. In the soft-photon approximation, 
the contribution of the virtual interference to the cross-section is 
\be
\label{if}
  d\sigma_{\mbox{\scriptsize virt}}(D_{if}) = 
      - d\sigma_{\mbox{\scriptsize Born}}\, \real\!
      \int \frac{d^{4}k}{(2\pi)^{4}}\, 
           \frac{32 i\pi\alpha\,(q_{1}\cdot k_{1})\,D_1}
                {[k^{2}-\lambda^2+io] [D_{1} - 2 (p_{1}\cdot k)] 
                 [-2(k_{1}\cdot k) + io] [-2(q_{1}\cdot k) + io]}.
\ee
Note that all particle poles are situated in the upper half of the complex 
$k_0$-plane. By closing the integration contour in the lower half-plane,
one finds that the complete virtual correction is equal to the photon-pole
contribution
\be
  d\sigma_{\mbox{\scriptsize virt}}(D_{if}) = 
      - d\sigma_{\mbox{\scriptsize Born}}\, \real\!
      \int \frac{d^{3}k}{(2\pi)^{3}\,2\omega}\,
           \frac{32\pi\alpha\,(q_{1}\cdot k_{1})\,D_1}
                {[D_{1} - 2 (p_{1}\cdot k)] [-2(k_{1}\cdot k) + io]
                 [-2(q_{1}\cdot k) + io]},
\ee
with $k_0 = \omega = \sqrt{\vec{k}^2+\lambda^2-io}$.
On the other hand, the corresponding bremsstrahlung interference can be 
written as 
\be
  d\sigma_{\mbox{\scriptsize real}}(D^{\mbox{\scriptsize R}}_{if}) = 
      - d\sigma_{\mbox{\scriptsize Born}}\, \real\!
      \int \frac{d^{3}k}{(2\pi)^{3}\,2\omega}\,
           \frac{32\pi\alpha\,(q_{1}\cdot k_{1})\,D_1^*}
                {[D_{1}^* + 2 (p_{1\cdot} k)] [2(k_{1}\cdot k) - io]
                 [-2(q_{1}\cdot k) + io]}.
\ee
By comparing the last two expressions, one can readily derive that the virtual 
and real interferences only differ by an overall minus sign and the 
substitution $D_{1}\to-D_{1}^{*}$. In the next section we derive an explicit 
expression for infrared-divergent virtual scalar four-point functions [see
Eq.(\ref{eq:feynman/IRD/D0134})].%
\footnote{In our example we need Eq.\,(\protect\ref{eq:feynman/IRD/D0134})
with $k_{2}$ replaced by $-q_1$. Note that, as a result of this
substitution, the invariant $s_{12}$ becomes negative.} From this 
expression one can see that the substitution 
$D_{1}\to-D_{1}^{*}$ does not change the real part of the interference  
(\ref{if}), i.e.~the sum of virtual and real interferences gives rise to a 
vanishing non-factorizable correction. Analogously, no other non-factorizable 
initial--final and initial--intermediate state photonic interferences 
contribute to the double resonant cross-section, if both virtual and real 
corrections are included. Similar arguments can be used to prove that 
initial-state up--down QED interferences vanish in our approach.

\section{Modified standard technique: \\Feynman-parameter integrals}
\label{sec:feynman}

In this section we present the calculation of the relevant virtual scalar
functions, using Feynman-parameter integrals. In addition the photon-pole 
parts of these functions are given, from which the real-photon corrections can 
be extracted. The striking difference with the usual calculations of scalar 
integrals lies in the systematic application of the soft-photon approximation.

\subsection{Scalar four-point functions in the soft-photon approximation}
\label{sec:feynman/general}

In this subsection we illustrate how to calculate a virtual scalar 
four-point function in the soft-photon approximation and how to extract the
photon-pole part. Consider to this end
\be
\label{eq:feynman/general/def_virt}
  D_{\mbox{\scriptsize virt}} = \int \frac{d^{4}k}{(2\pi)^{4}}\,
       \frac{1}{[k^2 + io] [2(p_{1}\cdot k) + D_{1} + io]
                [2(p_{2}\cdot k) + D_{2} + io] [2(p_{3}\cdot k) + D_{3} + io]},
\ee
where $D_{i}=p_{i}^{2}-M_{i}^{2}$. In general, the energy components $p_i^0$ of
the arbitrary momenta $p_i$  are not necessarily positive. In contrast to
the usual Feynman-parameter technique, where the Feynman-parameter 
transformation is applied to all propagators, we apply it only to propagators 
that are linear in $k$: 
\be
  D_{\mbox{\scriptsize virt}} = 2 \int \limits_{0}^{1}
       d^{3}\xi\, \delta\biggl( 1-\sum_{i=1}^3 \xi_{i} \biggr)\,
       I_{\mbox{\scriptsize virt}}(\xi),
\ee
with
\be
  I_{\mbox{\scriptsize virt}}(\xi) = \int \frac{d^{4}k}{(2\pi)^{4}}\, 
       \frac{1}{[k_0^{2} - \vec{k}^{2} + io] 
                [k_0 E(\xi) - \vec{k}\!\cdot\!\pvec(\xi) + A(\xi) + io]^{3}}.
\ee
The quantities $A(\xi)$ and $p^{\mu}(\xi)$ are given by
\be
  A(\xi) = \sum_{i=1}^3 \xi_i \, D_i \ \ \ \mbox{and}\ \ \ 
  p^{\mu}(\xi) = 2\sum_{i=1}^3 \xi_i \,p_i^{\mu}.
\ee
The energy component $E(\xi)$ of $p^{\mu}(\xi)$ can be positive or negative.
However, there is a freedom to choose $E(\xi)\leq 0$, because one can always 
perform a transformation of variables $k_0 \to -k_0$. Then 
\be
  I_{\mbox{\scriptsize virt}}(\xi) = \int \frac{d^{4}k}{(2\pi)^{4}}\, 
      \frac{1}{[k_0^{2} - \vec{k}^{2} + io] 
             [-k_0 |E(\xi)| - \vec{k}\!\cdot\!\pvec(\xi) + A(\xi) + io]^{3}}.
\ee
In the complex $k_0$-plane the denominators give rise to poles.
There are two photon poles, one in the upper and one in the lower half-plane.
The second denominator gives rise to a ``particle'' pole in the upper 
half-plane, for any value of $\xi_i$. It should be noted that this combines 
the 
three particle poles present in (\ref{eq:feynman/general/def_virt}), which 
could lie in the upper or lower half-plane. Closing the integration contour in 
the lower half-plane we get 
\be
\label{eq:feynman/general/1}
  I_{\mbox{\scriptsize virt}}(\xi) = - i \int \frac{d^{3}k}{(2\pi)^{3}}\, 
      \frac{1}{2\, |\vec{k}|\, 
      (-|\vec{k}|\,|E(\xi)| - x |\vec{k}|\,|\pvec(\xi)| + A(\xi) + io)^{3}},
\ee
where $x=\cos\theta$, with $\theta$ being the angle between $\pvec(\xi)$ and 
$\vec{k}$. It is not very difficult to perform the rest of the integrations in
momentum space. The final result is
\be
\label{eq:feynman/general/virtual}
  I_{\mbox{\scriptsize virt}}(\xi) = - \frac{i}{8\pi^{2}}\, 
     \frac{1}{A(\xi)\, (p^{2}(\xi) - io |E(\xi)|)}.
\ee

As we have seen in the previous section, the real-photon radiative 
interferences can be obtained from the photon-pole parts of the virtual 
corrections. Let us therefore consider the photon-pole part of 
(\ref{eq:feynman/general/def_virt}) in the lower half of the 
complex $k_0$-plane:
\be
  D_{\mbox{\scriptsize virt}}^{\gamma} = 
       - i \int \frac{d^{3}k}{(2\pi)^{3}}\,\frac{1}{2\, |\vec{k}|\, 
       [2(p_{1}\cdot k) + D_{1} + io] [2(p_{2}\cdot k) + D_{2} + io] 
       [2(p_{3}\cdot k) + D_{3} + io]},
\ee
with $k_0=\sqrt{\vec{k}^2-io}$. One can again proceed by introducing the 
Feynman parameters to obtain 
\be
\label{eq:feynman/general/2}
  I_{\mbox{\scriptsize virt}}^{\gamma}(\xi) = 
     - i \int \frac{d^{3}k}{(2\pi)^{3}}\, \frac{1}{2\,|\vec{k}|\,
     \Bigl[ E(\xi)\,\sqrt{\vec{k}^2-io} - x\,|\vec{k}|\,|\pvec(\xi)| 
            + A(\xi) + io \Bigr]^{3}}.
\ee
Equations (\ref{eq:feynman/general/1}) and (\ref{eq:feynman/general/2}) are
the same up to small modifications. In the case of the full virtual scalar 
function, Eq.\,(\ref{eq:feynman/general/1}) was obtained after contour 
integration in the complex $k_0$-plane. In that case we had the freedom to 
choose the contour in such a way that $E(\xi)\leq 0$. Now we have no such 
freedom. So, $E(\xi)$ cannot be considered as a negative quantity any more. 
It is clear that the final answer will be
\be
\label{eq:feynman/general/real}
  I_{\mbox{\scriptsize virt}}^{\gamma}(\xi) = 
       - \frac{i}{8\pi^{2}}\, 
       \frac{1}{A(\xi)\, [p^{2}(\xi) + io E(\xi)]}.
\ee
This expression is very similar to the one derived for the full virtual scalar
function. It can in fact be rewritten as 
\be
\label{eq:feynman/general/virt-real}
  I_{\mbox{\scriptsize virt}}^{\gamma}(\xi) = 
       - \frac{i}{8\pi^{2}}\,\frac{1}{A(\xi)}\, 
      \Biggl\{ \frac{1}{p^{2}(\xi) - io |E(\xi)|}
      - 2 i \pi\, \theta[E(\xi)]\,\delta[p^{2}(\xi)]
      \Biggr\},
\ee
where the first term in the curly brackets corresponds to the full virtual 
scalar function and the second term is the necessary modification. 
The second term in Eq.\,(\ref{eq:feynman/general/virt-real}) is the analogue of
the ``particle''-pole contribution in the approach of \cite{melyak}. Note that 
this term has an extra factor $i$. If all quantities were to be real 
(stable-particle case), then this term would not contribute to the 
non-factorizable correction to the cross-section, for which only the real part 
is important. In the case of unstable particles, this ``particle''-pole 
contribution is felt by the imaginary parts of the $W$-boson propagators, 
resulting in a potentially non-zero contribution to the cross-section. 
If one were to evaluate the photon-pole part of 
(\ref{eq:feynman/general/def_virt}) in the upper half of the complex 
$k_0$-plane, one
merely would have to replace $E(\xi)$ by $-E(\xi)$ in 
Eqs.\,(\ref{eq:feynman/general/real}) and (\ref{eq:feynman/general/virt-real}).

In practice, we calculate the relevant four-point functions 
$D_{\mbox{\scriptsize virt}}$ as well as the corresponding particle-pole 
contributions $D_{\mbox{\scriptsize virt}}^{\mbox{\scriptsize part}}$. The 
photon-pole part $D_{\mbox{\scriptsize virt}}^{\gamma}$ is obtained as 
$D_{\mbox{\scriptsize virt}}^{\gamma} = D_{\mbox{\scriptsize virt}}
- D_{\mbox{\scriptsize virt}}^{\mbox{\scriptsize part}}$, which can then be 
used to evaluate the real-photon radiative interferences. The complex 
half-plane where the particle-pole (photon-pole) contributions should be 
evaluated 
is fixed according to the rules given in 
Sect.\,\ref{sec:feynman/prescriptions}.

\subsection{Calculation of the virtual scalar four-point functions}

In this subsection we present the calculation of the virtual scalar four-point
functions and the associated photon-pole parts. Everything is considered 
in the 
soft-photon approximation. Since the four-point function $D_{1234}$ does not 
involve 
this approximation, we defer the corresponding results to 
App.\,\ref{app:feynman/D1234} 
and merely refer to the literature \cite{denner,gj} for its derivation.

Before listing the various results, we define our notation. To write 
down the analytical results we need to introduce some kinematic invariants:
\begin{eqnarray}
\label{invariants}
  && m_{1,2}^2 = k_{1,2}^2, \ \ \ 
     s = (p_1 + p_2)^2, \ \ \ 
     s_{12}  = ( k_1 + k_2)^2, \nonumber \\[1mm] 
  && s_{211^{\prime}} = ( k_2 + k_1 + k^{\prime}_1)^2, \ \ \ 
     s_{122^{\prime}} = ( k_1 + k_2 + k^{\prime}_2)^2,
\end{eqnarray}
and some short-hand notations:
\begin{eqnarray}
\label{shorthand}
  && y_0 = \frac{D_1}{D_2}, \ \ \  
     x_s = \frac{\beta-1}{\beta+1} + io, \ \ \ 
     \beta = \sqrt{1-4M_W^2/s} \, , \nonumber \\[1mm]
  && \zeta = 1 - \frac{s_{122^{\prime}}}{M_W^2} - io, \ \ \ 
     \zeta^{\prime} = 1 - \frac{s_{211^{\prime}}}{M_W^2} - io\, . 
\end{eqnarray}

\subsubsection{The virtual infrared-finite four-point function}
\label{sec:feynman/IRF}

We start off with the calculation of the infrared-finite scalar four-point 
function $D_{0123}$, which corresponds to the first diagram shown in 
Fig.~\ref{fig:1}. This function is infrared-finite owing to the presence of 
finite decay widths in the propagators of the unstable $W$ bosons. In the 
soft-photon limit we find
\be
  D_{0123} = \int \frac{d^{4}k}{(2\pi)^{4}}\, 
             \frac{1}{[k^2 + i o] [D_{1} - 2(p_{1}\cdot k)] 
                      [D_{2} + 2(p_{2}\cdot k)] [-2(k_{1}\cdot k) + io]},
\ee
where $D_{1,2}=p_{1,2}^{2}-M_{W}^{2}+io$. Originally the quantities $D_{1,2}$ 
are real, with the usual infinitesimal imaginary part. At the end of the 
calculation the analytical continuation to finite imaginary parts can be 
performed. Then $D_{1,2}=p_{1,2}^{2}-M_{W}^{2}+iM_{W}\Gamma_{W}$. 

Applying the Feynman-parameter technique as explained in 
Sect.\,\ref{sec:feynman/general}, we obtain the following representation 
\be
\label{eq:feynman/IRF/appendix}
  D_{0123} = \frac{- i}{4 \pi^{2}} \int\limits_0^1 d^{3}\xi\,
             \delta \biggl(1 - \sum \limits_{i=1}^3\xi_{i}\biggr)\ 
             \frac{1}{A(\xi)\,[p^{2}(\xi) - io]}
\ee
with
\be
  A(\xi) = \xi_{1}\,D_{1} + \xi_{2}\,D_{2}, \ \ \ 
  p^{\mu}(\xi) = - 2\xi_{1}\, p_{1}^{\mu} + 2\xi_{2}\, p_{2}^{\mu}
                 - 2\xi_{3}\, k_{1}^{\mu}.
\ee

As was indicated before, the integral will be calculated for small final-state
fermion masses and in the double-pole approximation. This implies 
\be
  2(p_{1}\cdot k_{1}) \approx p_{1}^{2} \approx M_W^2 \ \ \ \mbox{and}\ \ \ 
  2(p_{2}\cdot k_{2}) \approx p_{2}^{2} \approx M_W^2.
\ee

What is left is the integral over the space of Feynman parameters. The details 
of the integration are presented in App.\,\ref{app:feynman/IRF}. 
The final answer reads
\begin{eqnarray}
\label{eq:feynman/IRF/D0123}
  D_{0123} &=& \frac{i}{16\pi^2 M_W^2}\,\frac{1}{[D_2 - \zeta D_1]}
                      \Biggl\{
                          2\,\li\biggl(\frac{1}{y_0};\frac{1}{\zeta}\biggr)
                          - \li\biggl(x_s;\frac{1}{y_0}\biggr)
                          - \li\biggl(\frac{1}{x_s};\frac{1}{y_0}\biggr)
                      \Biggr. \nonumber \\[1mm]
           & &        \Biggl. {}+ \li\biggl(x_s;\zeta\biggr) 
                          + \li\biggl(\frac{1}{x_s};\zeta\biggr)
                          + \biggl[ \ln\biggl(\frac{M_W^2}{m_1^2}\biggr)
                                    \!+\! 2\ln(\zeta) \biggr]
                            \biggl[ \ln(y_0) \!+\! \ln(\zeta)\biggr] 
                      \Biggr\}.
\end{eqnarray}
The function $\li(x;y)$ is the continued dilogarithm
\be
  \li(x;y) = \Li(1-xy) + \ln(1-xy)\,[\ln(xy)-\ln(x)-\ln(y)],
\ee
with $\Li(x)$ the usual dilogarithm and $x,y$ lying on the first Riemann sheet.
The answer for the second infrared-finite scalar four-point function, 
$D_{0124}$, can be obtained from Eq.\,(\ref{eq:feynman/IRF/D0123}) by 
substituting $(p_1,k_1) \leftrightarrow (p_2,k_2)$.

\subsubsection{Photon-pole part of the infrared-finite four-point function}
\label{sec:feynman/IRFreal}

\begin{figure}[t]
  \unitlength 1cm
  \begin{center}
  \begin{picture}(13.4,7)
  \put(11,0.8){\makebox[0pt][c]{\boldmath $\xi_{2}$}} 
  \put(3.5,7){\makebox[0pt][c]{\boldmath $\xi_{1}$}}  
  \put(9,0.8){\makebox[0pt][c]{\boldmath $1$}} 
  \put(3.5,5.7){\makebox[0pt][c]{\boldmath $1$}}
  \put(7.9,0.7){\makebox[0pt][c]{\boldmath $\frac{1+\beta}{2}$}}  
  \put(6.6,0.8){\makebox[0pt][c]{\boldmath $\xi_{2}^{*}$}} 
  \put(3.3,4.65){\makebox[0pt][c]{\boldmath $\frac{1+\beta}{2}$}} 
  \put(3.3,2.13){\makebox[0pt][c]{\boldmath $\frac{1-\beta}{2}$}}
  \put(7.8,6.2){\makebox[0pt][c]{\boldmath $\downarrow$}}
  \put(7.5,6.9){\makebox[0pt][l]{\boldmath 
                $\xi_1 = \frac{\xi_2(s+s_{122^{\prime}}) - s_{122^{\prime}}}
                              {s-s_{122^{\prime}}}$}}
  \put(2,-2){\includegraphics{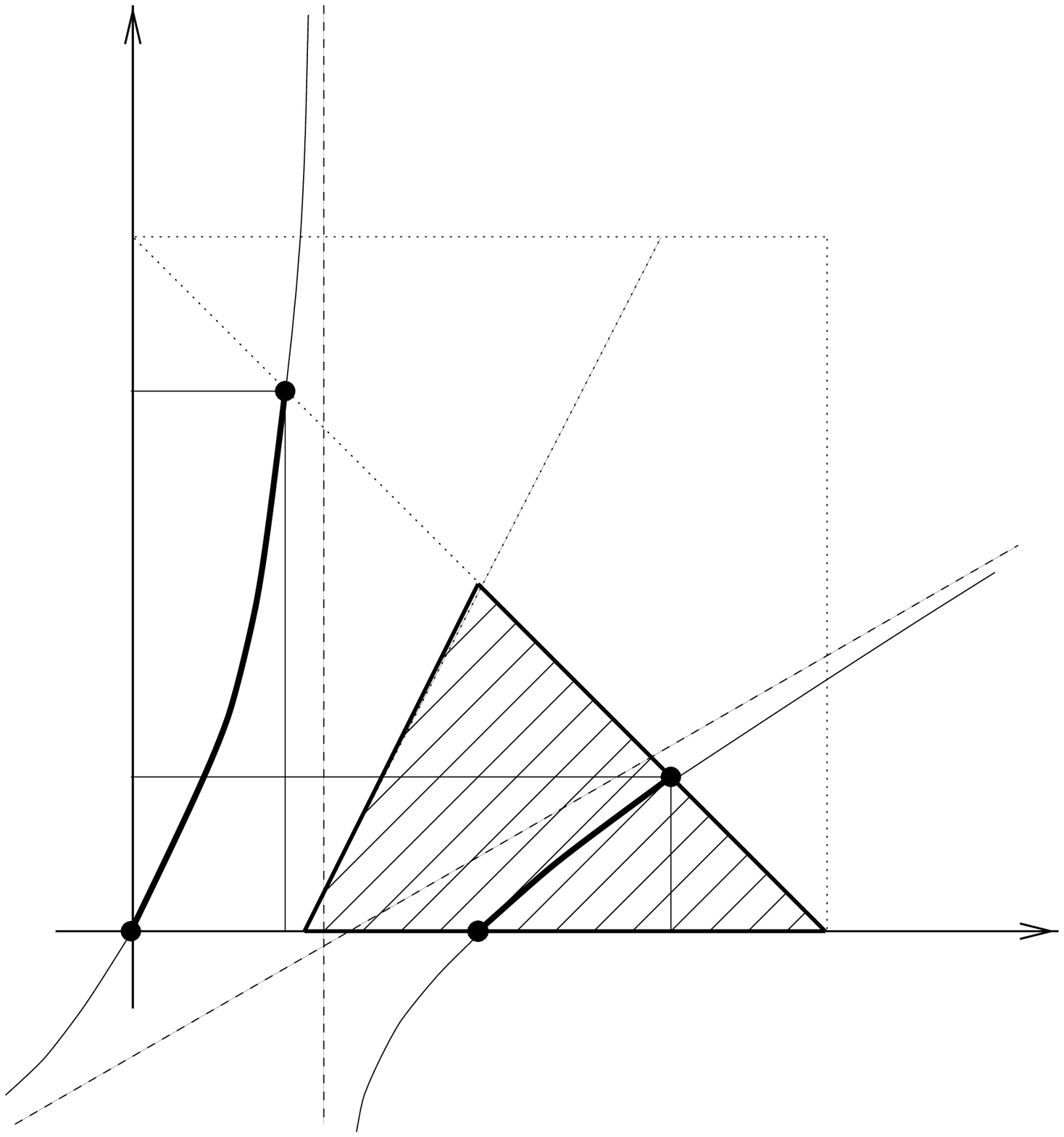}}
  \end{picture}
  \end{center}
  \caption[]{The integration area (shaded region) in the $(\xi_1,\xi_2)$ 
             Feynman-parameter space for the calculation of the 
             particle-pole part $D_{0123}^{\mbox{\scriptsize part}}$ of the
             infrared-finite scalar four-point function $D_{0123}$. The 
             thicker curves indicate the solutions of $p^2(\xi)=0$, with 
             $\xi_{2}^{*}=1-M_{W}^{2}/s_{122^{\prime}}$.}
\label{fig:feynman/IRFreal}
\end{figure}

As was explained in Sect.\,\ref{sec:feynman/general}, the photon-pole part of 
a scalar function can be obtained from the full scalar function by subtracting 
the ``particle''-pole contributions. According to 
Eq.\,(\ref{eq:feynman/general/virt-real}), the ``particle''-pole 
contributions in the lower half of the complex $k_0$-plane are given by
\be
  D_{0123}^{\mbox{\scriptsize part}} = \frac{1}{2 \pi}
         \int \limits_{0}^{\xi_{1} + \xi_{2} < 1}
         \frac{ d \xi_1\,d \xi_2}{A(\xi)}\,\theta[E(\xi)]\,\delta[p^{2}(\xi)].
\ee
The integration area is defined by the $\theta$-function for the energy and by 
the condition $\xi_{1}+\xi_{2}<1$. The allowed area of integration and the 
curve where the $\delta$-function has a non-zero value are schematically shown 
in Fig.\,\ref{fig:feynman/IRFreal}. The depicted situation represents the
most general case for the kinematics we are interested in.

After the integration over the $\delta$-function has been performed, one is 
left with a simple one-dimensional integration of logarithmic type. The final 
result is
\be
  D_{0123}^{\mbox{\scriptsize part}} = 
        \frac{1}{8 \pi M_W^2}\,\frac{1}{D_2 - \zeta D_1}
        \Biggl[ \ln\bigl( 1 - y_0 x_s \bigr) - \ln\bigl( 1 - x_s/\zeta \bigr)
        \Biggr].
\ee

\subsubsection{The virtual infrared-divergent four-point function}
\label{sec:feynman/IRD}

In this subsection we describe the calculation of the infrared-divergent  
scalar four-point function $D_{0134}$, which enters the non-factorizable
corrections through the decomposition of the five-point function.
Similar four-point functions may also appear in the initial--final state 
interactions, but, as was mentioned before, the interactions of this type 
vanish in the sum of virtual and real contributions. The four-point function
\be
          D_{0134} = \int \frac{d^{4}k}{(2\pi)^{4}}\,\frac{1}
                   {[k^2 - \lambda^{2} + io] [D_{1} - 2 (p_{1}\cdot k)]
                    [-2(k_{1}\cdot k) + io] [ 2 (k_{2}\cdot k) +io ]}.
\ee
is infrared-divergent, because only one unstable particle is involved, which 
is not enough to regularize the divergence. Therefore we introduce a regulator
mass $\lambda$ for the photon in order to trace the cancellation of infrared 
divergences in virtual and real corrections. Again we can apply the 
Feynman-parameter technique as explained in Sect.\,\ref{sec:feynman/general}.
However, special care has to be taken with the photon mass $\lambda$.
As usual we can introduce Feynman parameters according to  
\be
\label{eq:feynman/IRD/feynman_parameters}
  D_{0134} = 2 \int \limits_{0}^{1} d^{3}\xi \, 
        \delta\biggl(1 - \sum_{i=1}^3 \xi_{i}\biggr) \, I_{0134}(\xi),
\ee
with
\be
\label{eq:feynman/IRD/appendix}
  I_{0134}(\xi) = \int \frac{d^{4}k}{(2 \pi)^{4}}\,
         \frac{1}{[k^{2} - \lambda^{2} + i o] 
         [- k_0\,|E(\xi)| - \vec{k}\!\cdot\!\pvec(\xi) + A(\xi) + i o ]^{3}}
\ee
and 
\be
  A(\xi) = \xi_{2}\,D_{1}, \ \ \ 
  p^{\mu}(\xi) = - 2\xi_{1}\, k_{1}^{\mu} - 2\xi_{2}\, p_{1}^{\mu}
                 + 2\xi_{3}\, k_{2}^{\mu}.
\ee
Again we can exploit the freedom to perform the variable transformation 
$k_0 \to -k_0$ in order to fix the sign of the energy component $E(\xi)$.
After the integration over momentum space, the details of which can be found
in App.\,\ref{app:feynman/IRD}, we obtain
\be
\label{eq:feynman/IRD/1}
  I_{0134}(\xi) = - \frac{i}{8 \pi^{2}}\, 
          \frac{\partial}{\partial p^{2}}\,
          \biggl\{
          \frac{1}{\sqrt{A^{2} - \lambda^{2} p^{2}}}
          \ln \biggl( \frac{A - \sqrt{A^{2} - \lambda^{2} p^{2} }}
                           {A + \sqrt{A^{2} - \lambda^{2} p^{2} }}\,
              \biggr)
          \biggr\}                             
\ee
with
\be
  \frac{p^{2}(\xi)}{4} = \xi_{2}^{2} M_{W}^{2} + \xi_{1}^{2} m_{1}^{2} 
      + \xi_{3}^{2} m_{2}^{2} + \xi_{1} \xi_{2} M_{W}^{2} 
      - \xi_{1} \xi_{3} s_{12} + \xi_{2} \xi_{3} 
   (M_{W}^{2} - s_{211^{\prime}}).
\ee
The masses of the final-state fermions, $m_{1}$ and $m_{2}$, are taken to be 
small in our approximation.

One is left with a twofold Feynman-parameter integration 
(see App.\,\ref{app:feynman/IRD}), which results in
\begin{eqnarray}
\label{eq:feynman/IRD/D0134}
  D_{0134} &=& - \frac{i}{16 \pi^2 s_{12}}\,\frac{1}{D_1}
               \Biggl[ \Li\biggl(1 + \frac{\zeta^{\prime} M_W^2}{s_{12}}\biggr)
                       - 2 \ln \biggl( \frac{M_W \lambda}{-D_1} \biggr)
                           \ln \biggl( \frac{m_1 m_2}{-s_{12}-io} \biggr)
               \Biggr. \nonumber \\[1mm]
           & & \Biggl. \hphantom{- \frac{i}{16 \pi^2 s_{12}}\,\frac{1}{D_2}a}
                       + \frac{\pi^2}{3}
                       + \ln^2\biggl( \frac{M_W}{m_1} \biggr)
                       + \ln^2\biggl( \frac{m_2}{\zeta^{\prime} M_W} \biggr)
               \Biggr].
\end{eqnarray}
The answer for the second infrared-divergent scalar four-point function, 
$D_{0234}$, can be obtained from Eq.\,(\ref{eq:feynman/IRD/D0134}) by 
substituting $(p_1,k_1) \leftrightarrow (p_2,k_2)$.

\subsubsection{Photon-pole part of the infrared-divergent four-point function}
\label{sec:feynman/IRDreal}

\begin{figure}[t]
  \unitlength 1cm
  \begin{center}
  \begin{picture}(13.4,7)
  \put(4.5,7.5){\makebox[0pt][c]{\boldmath $t$}} 
  \put(11.6,2.4){\makebox[0pt][c]{\boldmath $u$}} 
  \put(8.1,2.3){\makebox[0pt][c]{\boldmath $u^{*}$}}
  \put(6.6,3.0){\makebox[0pt][c]{\boldmath $\frac{s}{s_{211^{\prime}}}$}}
  \put(5.7,3.0){\makebox[0pt][c]{\boldmath $\frac{-1}{\zeta^{\prime}}$}}
  \put(11,7){\makebox[0pt][c]{\small $E(u,t)=0$}}  
  \put(10.8,6.5){\makebox[0pt][c]{\boldmath $\downarrow$}}
  \put(2.5,0.4){\makebox[0pt][c]{\small $p^{2}<0$}} 
  \put(9,0.4){\makebox[0pt][c]{\small $p^{2}>0$}}
  \put(9,6){\makebox[0pt][c]{\small $p^{2}<0$}}
  \put(-10.3,-14.5){\includegraphics{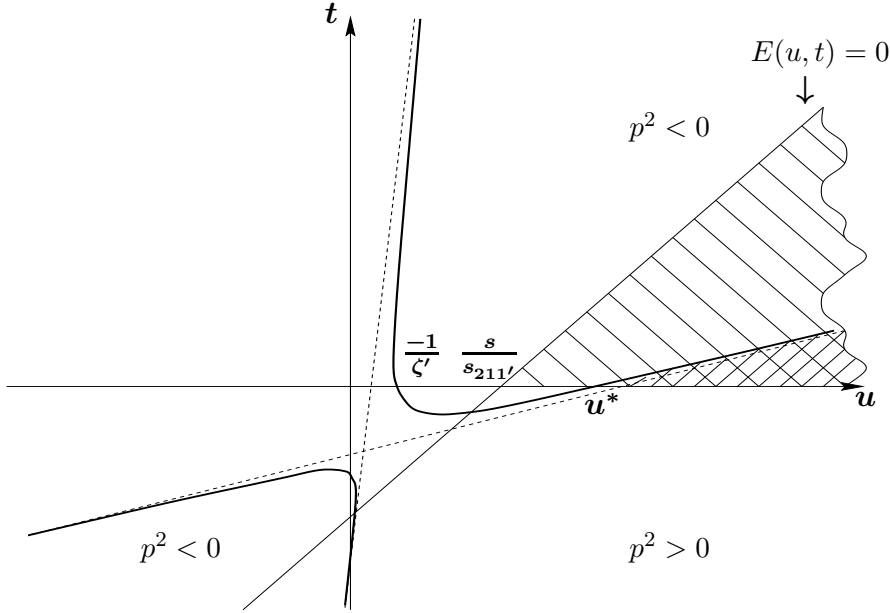}}
  \end{picture}
  \end{center}
  \caption[]{The area of integration in the $(u,t)$-plane for the calculation 
of 
             the particle pole $D_{0134}^{\mbox{\scriptsize part}}$.
             The shaded region is the area of integration where $E(u,t)>0$.
             The doubly-shaded region is the area of integration 
where $E(u,t)>0$ 
	     and $p^{2}(u,t)>0$.
             The thicker curves indicate the solutions of $p^{2}(u,t)=0$, with 
             $u^{*}=-\zeta^{\prime}M_{W}^{2}/m_{2}^{2}$.}
\label{fig:feynman/IRDreal}
\end{figure}

The procedure for calculating the photon-pole part $D_{0134}^{\gamma}$ follows
our general strategy, i.e.~we calculate the corresponding ``particle''-pole 
contributions and subtract them from the full virtual scalar function.
The difference between this case and the one discussed in 
Sect.\,\ref{sec:feynman/IRFreal} lies in the fact that here both 
photon-pole and ``particle''-pole contributions are infrared-divergent. 
So as to keep track of the cancellation of the infrared divergences, we 
again introduce a regulator mass $\lambda$ for the photon.

In App.\,\ref{app:feynman/IRD} the following convenient representation for
$I_{0134}(\xi)$ was derived in Eq.\,(\ref{eq:app:feynman/IRD}):
\be
          I_{0134}(\xi) = \frac{i}{16 \pi^{2} |E(\xi)|} \,
\int \limits_{-\infty}^{\infty} 
          \frac{d z }{\Bigl[-\, |E(\xi)| \,\sqrt{z^{2} + \lambda^{2}} 
                            - |\vec{\, p}(\xi)| \,z + A(\xi) + i o \Bigr]^{2}}.
\ee
{}From the general discussion in Sect.\,\ref{sec:feynman/general} we know that
the photon-pole contribution to the four-point functions in the 
Feynman-parameter 
representation, $I_{\mbox{\scriptsize virt}}^{\gamma}(\xi)$, can be obtained 
from the complete 
virtual function, $I_{\mbox{\scriptsize virt}}(\xi)$, by a substitution 
$|E(\xi)|\to-E(\xi)$.
Then the ``particle''-pole contribution, 
$D_{0134}^{\mbox{\scriptsize part}}=D_{0134}-D_{0134}^{\gamma}$,
has the form
\be
        D_{0134}^{\mbox{\scriptsize part}} = 2 \int \limits_{0}^{1} 
d^{3}\xi \, 
        \delta\biggl(1 - \sum_{i=1}^3 \xi_{i}\biggr) \, 
        I_{0134}^{\mbox{\scriptsize part}}(\xi),
\ee
where 
\be
          I^{\mbox{\scriptsize part}}_{0134}(\xi) = 
	-\frac{i \theta[E(\xi)]}{8\pi^{2}E(\xi)}\,
\frac{\partial}{\partial E(\xi)}
	\int\limits_{-\infty}^{+\infty}
	\frac{E(\xi)\, dz}{[E(\xi)]^{2}\,
(z^{2}+\lambda^{2})-(|\pvec(\xi)|\,z-A(\xi)-io)^{2}}.
\ee
By analysing the pole structure of the last expression, it is easy to realize 
that $I_{0134}^{\mbox{\scriptsize part}}(\xi)$ is only non-vanishing for 
$p^{2}(\xi)\geq 0$.
This results in a simple formula
\be
       I^{\mbox{\scriptsize part}}_{0134}(\xi) = 
       \frac{\theta[E(\xi)]}{4\pi}\,\frac{\partial}{\partial p^{2}(\xi)}\,
	\left\{
	\frac{\theta[p^{2}(\xi)]}
             {A(\xi)\sqrt{1-\frac{\lambda^{2}p^{2}(\xi)}{[A(\xi)+io]^{2}}}}
	\right\}.
\ee
What is left is the integration over the space of Feynman parameters.
In order to simplify the calculation it is advisable to make the change of 
variables
$\,\xi_{1}=t/(1+t+u)\,$ and $\,\xi_{2}=1/(1+t+u)$. The area of integration 
in the $(u,t)$-plane is shown schematically in Fig.\,\ref{fig:feynman/IRDreal}.
The final result is
\begin{equation}
\label{eq:IRD/part}
        D_{0134}^{\mbox{\scriptsize part}} = 
	\frac{1}{8 \pi s_{12} D_{1}}
	\Biggl[
 	-\ln(-\zeta^{\prime}) 
        + \ln\biggl(\frac{D_{1}}{i M_{W}^{2}}\biggr) 
	- \ln\biggl(\frac{\lambda}{m_{2}}\biggr)
	\Biggr].
\end{equation}
This is the same result as obtained in Sect.\,\ref{sec:dmi}, where we will 
approach the various calculations in a significantly different way.

\subsection{The Coulomb-like scalar three-point function}
\label{sec:feynman/coulomb}

As was pointed out in Sect.\,\ref{sec:gaugeinv}, a gauge-invariant definition 
of the non-factorizable corrections requires the proper inclusion of a 
Coulomb-like contribution. In this subsection we calculate the associated
scalar three-point function in the soft-photon approximation. This three-point 
function is infrared-finite, but ultraviolet-divergent. This divergence occurs 
as a result of the fact that we neglect the $k^{2}$ dependence of the 
propagators, following our general soft-photon strategy. Although the virtual 
and real Coulomb-like contributions to the cross-section are separately 
ultraviolet-divergent, the sum is finite.

The virtual Coulomb-like scalar three-point function $C_{012}$ is defined as
(see Fig.~\ref{fig:2}):
\begin{equation}
          C_{012} = \int\frac{d^{4}k}{(2\pi)^{4}}
          \frac{1}{[k^{2}+io] [D_{1}-2(p_{1}\cdot k)] [D_{2}+2(p_{2}\cdot k)]}.
\end{equation}
Similar to the calculation of the scalar four-point functions, we introduce 
the Feynman parameters only for the propagators that are linear in $k$.
We limit the area of integration over $\vec{k}$ by the condition 
$|\vec{k}|<\Lambda$, where $\Lambda \gg \Gamma_{W}$. After having performed 
the momentum integration, one is left with a one-dimensional integral over 
Feynman parameters:
\begin{eqnarray}
  C_{012} &=& -\frac{i}{8\pi^{2}}\int\limits_0^1 d^{2}\xi\,
              \delta(1-\xi_{1}-\xi_{2})\,
              \Biggl[
              \frac{1}{|\pvec|\,(|E|-|\pvec|)}\,
                  \ln\biggl( \frac{|E|-|\pvec|-A/\Lambda}{-A/\Lambda} \biggr)
              \Biggr. \nonumber \\[1mm]
          & & \Biggl.
            {}- \frac{1}{|\pvec|\,(|E|+|\pvec|)}\,
                  \ln\biggl( \frac{|E|+|\pvec|-A/\Lambda}{-A/\Lambda} \biggr)
              \Biggr],
\end{eqnarray}
where $A(\xi) = \xi_{1}\,D_{1} + \xi_{2}\,D_{2}$ and 
$p^{\mu}(\xi) = -2\xi_{1}\, p_{1}^{\mu} + 2\xi_{2}\, p_{2}^{\mu}$.
The final integration over the Feynman parameters yields
\begin{eqnarray}
\label{C012}
  C_{012} &=& \frac{i}{16\pi^2 s \beta} \Biggl\{ 
                \li\biggl( y_0;\frac{1}{x_s} \biggr)
              + \li\biggl( \frac{1}{y_0};\frac{1}{x_s} \biggr) 
              - 2\,\Li\biggl( 1-\frac{1}{x_s} \biggr) 
              + \frac{1}{2}\,\ln^2(y_0) \Biggr. \nonumber \\[1mm]
          & &                           \Biggl. {}
              \hphantom{\frac{i}{16\pi^2 s \beta}a}
              + \ln(x_s) \Biggl[ \ln\biggl( \frac{-iD_1}{2M_W \Lambda} \biggr)
                + \ln\biggl( \frac{-iD_2}{2M_W \Lambda} \biggr) \Biggr]
              - 2i\pi\,\ln\biggl( \frac{1+x_s}{2} \biggr) 
                                        \Biggr\}.
\end{eqnarray}

In a similar way one can calculate the particle-pole contribution
$C_{012}^{\mbox{\scriptsize part}}$, which can be used to extract the 
photon-pole part $C_{012}^{\gamma}=C_{012}-C_{012}^{\mbox{\scriptsize part}}$.
The final answer for the particle-pole contribution reads
\be
  C_{012}^{\mbox{\scriptsize part}} = \frac{1}{8\pi s \beta} \Biggl\{
              \ln(1-x_s) + \ln(1+x_s) - \ln(1-y_0 x_s)
              - \ln\biggl( \frac{-iD_2}{M_W \Lambda} \biggr) \Biggr\}.
\ee
Note that the real part of $C_{012}^{\mbox{\scriptsize part}}$ does not 
contribute to the non-factorizable corrections. Therefore, the $\Lambda$ 
dependence effectively drops out from the particle-pole contribution. 

One may wonder how the non-factorizable contributions (\ref{eq:mtrx_element}) 
and (\ref{Coulvirt}) to the cross-section
compare with the Coulomb contribution as calculated in the literature. Our
calculation is based on the assumption of being at least a few widths away 
from the 
threshold [the accuracy of this approximation is of
${\cal O}(\Gamma_{W}/\Delta E)$], whereas 
the Coulomb effect in the literature is valid in the 
non-relativistic region, where it approximates the cross-section with 
accuracy ${\cal O}(\beta)$. Therefore one could try to compare them
in an overlapping region, where $\Gamma_{W}\ll\Delta E\ll M_{W}$. The 
Coulomb effect
calculated in Ref.\,\cite{coulomb3} consists of two parts. One contribution
that is also present for on-shell $W$-bosons, and one that comes from the 
off-shellness. The former is related to factorizable corrections and the 
latter is related to non-factorizable corrections and vanishes upon 
integration 
over the 
virtualities\footnote{Only soft virtual photons contribute to both parts.
For the on-shell (factorizable) part of the Coulomb effect photons with 
momenta 
$\omega\approx \beta^{2}M_{W}$ and
$\,|\vec{k}|\,\approx \beta M_{W}$ are important (hence $k^{2}$ can not be 
neglected
in the propagators of the unstable particles). On the other hand, only photons 
with momenta $\omega\approx\Gamma_{W}$ and $|\vec{k}|\approx\Gamma_{W}/\beta$
give the leading contribution to the off-shell part of the Coulomb effect.
Far from the threshold, where $\Gamma_{W}\ll\Delta E\ll M_{W}$,
the two regions in the 
photon-momentum space are well separated. Because of this the effects 
are additive. Near threshold, where $\Gamma_{W}\approx\Delta E$, the two 
regions start to intersect.}.
In the overlap region $(\Gamma_{W}\ll\Delta E\ll M_{W})$ 
this off-shell term equals the $1/\beta$ part of our full expression.

\section{Direct momentum-integration method}
\label{sec:dmi}

In this section we present an alternative method of calculating the 
non-factorizable corrections, i.e.~the so-called direct momentum-integration
(DMI) method. As in Sects.\,\ref {sec:decomp} and \ref{sec:feynman}, we use 
the soft-photon approximation and assume the charged final-state fermions 
to be 
massless, which is a good approximation for the process under consideration.
In contrast to Sects.\,\ref {sec:decomp} and \ref{sec:feynman} we do not make
any assumptions about the mass of the final-state neutrinos, because it does 
not 
simplify the calculation significantly. This gives us the opportunity to 
apply the
results of this section to top-quark pair production.
Following the approach of \cite{melyak}, we write the amplitudes corresponding 
to the diagrams shown in Fig.\,\ref{fig:1} in terms of virtual and real scalar 
four-/five-point functions. In contrast to the Feynman-parameter approach of 
Sect.\,\ref{sec:feynman}, we do not introduce Feynman parameters, but perform 
instead a direct integration over momentum space \cite{melyak}. The calculation
can be considerably simplified by an appropriate choice of the frame.

First we calculate the infrared-finite virtual and real four-point functions. 
The calculation is close to the one presented in \cite{melyak},
but in contrast to \cite{melyak} we make a clear separation between the virtual
and real contributions. Our final result 
agrees with the one of \cite{melyak}, as well as with the one obtained in the
MST in Sects.\,\ref{sec:feynman/IRF} and \ref{sec:feynman/IRFreal}. Next we
calculate the infrared-divergent virtual and real four-point functions. Again 
we perform a separation of the real and virtual contributions, and provide a 
careful treatment of the divergences. All this is needed in order to trace 
the cancellations of infrared and collinear divergences. We find complete 
agreement with our results obtained in Sects.\,\ref{sec:feynman/IRD} and 
\ref{sec:feynman/IRDreal}. At the same time the structure of the divergences 
in our results appears to be significantly different from the one obtained by
using the method of \cite{melyak}, even in the complete answer when virtual 
and real corrections are summed up. Although the infrared-divergent scalar
four-point functions do not appear directly in the answer for the 
non-factorizable corrections, the observed difference with the method of 
\cite{melyak} turns out to be indicative, because similar problems arise in 
the evaluation of the 
(infrared-divergent) five-point functions. Finally we calculate in the same way
the virtual and real five-point functions. After summing up the virtual and 
real five-point functions, we find, in contrast to the result in \cite{melyak},
that all collinear divergences cancel exactly, even for 
cross-sections that are exclusive with respect to the virtualities of the $W$
bosons.

In conclusion, the calculation presented in this section is an extension of 
the method of \cite{melyak}. We provide a proper treatment of the infrared and 
collinear divergences, and make a clear separation of the virtual and real 
corrections. Because of this, the calculation becomes much more involved.
We use the results obtained in this section as an independent check of the 
results of Sects.\,\ref{sec:decomp} and \ref{sec:feynman}. Although the methods
are completely different and the answer of this section is very 
complicated, a perfect 
numerical agreement between our two calculations is observed. 

Before listing the various results, we first define the notation. For the 
calculations in the DMI method we need to specify the momenta in the 
centre-of-mass frame of the initial state. Because of the soft-photon,
double-pole ($D_{1,2} \ll M_W^2$) approximation, the four-momenta of the two
intermediate $W$ bosons are related in a simple way:
\be
  p_1^{\mu} = (E,\pvec) = E\,(1,\vec{v}\,)\ \ \mbox{and}\ \ 
  p_2^{\mu} = (E,-\pvec) = E\,(1,-\vec{v}\,),
\ee
with $|\vec{v}| = \beta$ the (on-shell) velocity of the $W$ bosons [see also
Eq.\,(\ref{shorthand})]. The other relevant momenta are 
\be
  k_1^{\mu} = (E_1,\vec{k}_1) = E_1\,(1,\vec{v}_1)\ \ \mbox{and}\ \ 
  k_2^{\mu} = (E_2,\vec{k}_2) = E_2\,(1,\vec{v}_2),
\ee
with $|\vec{v}_i| \equiv v_i = \sqrt{1-m_i^2/E_i^2}\,$ for $(i=1,2)$. 
In addition we need the definition of some (polar) angles with respect to the 
direction of the $W^+$ boson: $\theta_i = \angle(\vec{v},\vec{v}_i)$ and 
$x_i = \cos\theta_i$ for $(i=1,2)$. The difference of the azimuthal angles of
$\vec{k}_1$ and $\vec{k}_2$ is given by $\phi_{12}$. So, for $\sin\phi_{12}=0$ 
the final-state three-momenta $\vec{k}_i$ and $\vec{k}^{\prime}_i$ lie in one
plane. In the plane spanned by $\vec{k}_1$ and $\vec{k}_2$ we define
$\theta_{12} = \angle(\vec{v}_1,\vec{v}_2)$ and $x_{12} = \cos\theta_{12}$.

\subsection{Non-factorizable infrared-finite corrections}
\label{sec:dmi/IRF}

In this subsection we briefly describe the calculation of the infrared-finite
four-point functions in the DMI scheme, following Ref.\,\cite{melyak}. The 
result agrees with the one presented in \cite{melyak}, so this subsection is 
merely presented for completeness. The contribution of the infrared-finite 
virtual four-point function $D_{0124}$ to the non-factorizable matrix element 
is given by
\be
  M_{0124} = \bar{M}_{B}\,\frac{i}{D_{1}}\int\frac{d^{4}k}{(2\pi)^{4}}\,
             \frac{16 \pi\alpha\,(p_{1}\cdot k_{2})}
                  {[k^{2}+io] [2(k\cdot k_{2})+io] [D_{1}-2(p_{1}\cdot k)]
                   [D_{2}+2(p_{2}\cdot k)]},       
\ee
where $M_{B}=\bar{M}_{B}/(D_1 D_2)$ is the Born matrix element of the 
process, involving the production of an intermediate $W$-boson pair and its 
subsequent decay. We start the calculation by decomposing the unstable
$W$-boson propagators according to
\be
\label{Wpropdecomp}
  \frac{1}{[D_{1}-2(p_{1}\cdot k)][D_{2}+2(p_{2}\cdot k)]}
  =
  \biggl[ \frac{1}{D_{1}-2(p_{1}\cdot k)} + \frac{1}{D_{2}+2(p_{2}\cdot k)}
  \biggr] \frac{1}{D+4\pvec\!\cdot\!\vec{k}}\, ,
\ee
where $D=D_{1}+D_{2}$. The first term has two particle poles: one in the lower 
and one in the upper half of the complex $k_0$-plane. We close the contour 
in the 
lower half-plane, resulting in one particle-pole and one photon-pole 
contribution. The second term has all its particle poles in the lower 
half-plane. By closing the integration contour in the upper half-plane, 
only one of 
the photon poles will contribute. Note that the above decomposition mixes 
photon- and particle-pole contributions. In order to avoid possible confusion 
with the pure photon- and particle-pole contributions, we will write 
$M_{0124}^{\mbox{\scriptsize`$\gamma$'}}$ and 
$M_{0124}^{\mbox{\scriptsize `part'}}$ if the decomposition is used.

\subsubsection{Particle-pole residue}

We first concentrate on the particle-pole residue contributing to the first 
term in Eq.\,(\ref{Wpropdecomp}). This particle pole is situated at 
$k_0 = \vec{v}_2\!\cdot\!\vec{k}$. The corresponding residue reads
\be
  M_{0124}^{\mbox{\scriptsize `part'}} = 
              \bar{M}_{B}\,\frac{4\pi\alpha}{D_{1}}\,
              \frac{1-\beta x_{2}}{2 E}
              \int \frac{d^{3}k}{(2\pi)^{3} [(\vec{v}_2\!\cdot\!\vec{k})^{2}
                                             -\vec{k}^{2}] }\,
              \frac{1}{\bigl[\frac{D}{2 E}+2\vec{v}\!\cdot\!\vec{k}\bigr]
              \bigl[\frac{D_{1}}{2 E}-(\vec{v}_{2}-\vec{v})\!\cdot\!\vec{k}
              \bigr]}.
\ee
The propagators can be exponentiated by introducing an integration over 
``time'':
\be
  M_{0124}^{\mbox{\scriptsize `part'}} = 
              - \bar{M}_{B}\,\frac{4\pi\alpha}{D_{1}}\,
              \frac{1-\beta x_{2}}{2 E}
              \int\limits_{0}^{\infty} d\tau\,d\tau_{1}\,
              e^{i\bigl[\frac{D}{2 E}\tau+\frac{D_{1}}{2 E}\tau_{1}\bigr]}
              \int\frac{d^{3}k}{(2\pi)^{3}}\,
              \frac{e^{i\vec{k}\cdot\vec{r}}}
                   {(\vec{v}_{2}\!\cdot\!\vec{k})^{2} - \vec{k}^{2}},
\ee
where 
\be
  \vec{r} = 2\tau\,\vec{v} - \tau_{1}\,(\vec{v}_{2}-\vec{v}).
\ee
The integral is infrared-finite, so there is no need to introduce 
a non-zero photon mass as infrared regulator. The spatial integration 
can be recognized 
as a relativistic Coulomb potential of a moving particle: 
\be
  \phi(r) = - 4\pi \int  \frac{d^{3}k}{(2\pi)^{3}}\,
            \frac{e^{i\vec{k}\cdot\vec{r}}}
                 {(\vec{v}_{2}\!\cdot\!\vec{k})^{2} - \vec{k}^{2}}
          = \frac{1}{\sqrt{r_{\parallel}^{2}+r_{\perp}^{2}(1-v_{2}^{2})}}.
\ee
Here $r_{\parallel}$ and $r_{\perp}$ are the absolute values  of the 
components of $\vec{r}$ parallel and perpendicular to $\vec{v}_2$:
\be
  r_{\parallel} = 2\beta x_{2} \tau - (1-\beta x_{2}) \tau_{1},\ \ \ 
  r_{\perp} = \beta (2 \tau + \tau_{1})\,\sin\theta_{2}.
\ee
Note that $1-v_{2}^{2}=m_{2}^{2}/E_{2}^{2}$ \,is small, but finite.

To do the remaining integrations over $\tau$ and $\tau_{1}$, we can make a 
change of variables according to $(\tau,\tau_{1})\to (\xi,y)$, 
with $\tau=\xi y$, $\tau_{1} = \xi (1-y)$, and the Jacobian 
$|\frac{\partial(\tau,\tau_{1})}{\partial(\xi,y)}|=\xi$.
The area of integration changes from $\tau>0$ and $\tau_{1}>0$ to 
$\xi>0$ and $0<y<1$. After this change of variables, the quantities
$r_{\parallel}$ and $r_{\perp}$ will be proportional to $\xi$, rendering 
the integration over $\xi$ trivial. The last integral over $y$ can be
calculated in a straightforward way, yielding after some manipulations
\begin{eqnarray}
\label{eq:dmi/IRF/part1}
  \underline{x_{2}<0}\!\!&:& M_{0124}^{\mbox{\scriptsize `part'}} = 
             \bar{M}_{B}\,i\alpha\,\frac{1-\beta x_{2}}{D_{1} \eta(x_{2})}
             \Biggl[ \ln\biggl( \frac{D}{D_{1}} \biggr) 
                   + \ln\biggl( \frac{1-\beta x_{2}}{-2 \beta x_{2}} \biggr)
             \Biggr], \\[1mm]
\label{eq:dmi/IRF/part2}
  \underline{x_{2}>0}\!\!&:& M_{0124}^{\mbox{\scriptsize `part'}} = 
             \bar{M}_{B}\,i\alpha\,\frac{1-\beta x_{2}}{D_{1} \eta(x_{2})}
             \Biggl[ \ln\biggl( \frac{\eta(x_{2})}{D}\biggr)
                   + \ln\biggl( \frac{\eta(x_{2})}{D_{1}}\biggr)  
                   + \ln\biggl( \frac{2 x_{2}(1-\beta x_{2})}
                                      {\beta (1-x_{2}^{2})} \biggr)
                   + \ln\biggl( \frac{E_{2}^{2}}{m_{2}^{2}} \biggr)
             \Biggr], \nonumber \\
\end{eqnarray}
where 
\be
  \eta(x) = (1 + x\beta)\,D_{1} + (1 - x \beta)\,D_{2}.           
\ee
The result for $M_{0124}^{\mbox{\scriptsize `part'}}$ is not the same for
$x_{2}<0$ and $x_{2}>0$. This is caused by the propagator decomposition
(\ref{Wpropdecomp}). However, the complete result, with the photon-pole 
residue 
included, will be independent of the sign of $x_{2}$.

\subsubsection{Photon-pole residues}

Now we calculate the photon-pole residues. There are two such contributions, 
one in each of the terms in the decomposition (\ref{Wpropdecomp}). We will
indicate these two contributions by 
$M_{0124}^{\mbox{\scriptsize`$\gamma$'},\,1}$ and
$M_{0124}^{\mbox{\scriptsize`$\gamma$'},\,2}$, respectively. For 
$M_{0124}^{\mbox{\scriptsize`$\gamma$'},\,1}$ the 
contour is closed in the lower half of the complex $k_0$-plane. In that 
case the 
photon pole is situated at $k_0 = |\vec{k}| - io$, yielding
\be
\label{MIRFphoton1}
  M_{0124}^{\mbox{\scriptsize`$\gamma$'},\,1} = \bar{M}_{B}\,
               \frac{\alpha}{2\pi}\,\frac{1-\beta x_{2}}{D_{1}}
               \int \frac{d^{2}\Omega_k}{2\pi}\,
               \frac{1}{\eta(x)\,(1-\vec{v}_{2}\!\cdot\!\vec{n}_{k})}\,
               \biggl[ \ln\biggl( \frac{2\beta x + i o}{1-\beta x} \biggr)
                     - \ln(D) + \ln(-D_{1})
               \biggr].
\ee
Here $\vec{n}_k$ stands for the unit vector in the $\vec{k}$ direction and 
$\Omega_k$ indicates the angular variables in spherical coordinates (with the
polar axis defined along $\pvec$). 
For $M_{0124}^{\mbox{\scriptsize`$\gamma$'},\,2}$ the 
contour is closed in the upper half of the complex 
$k_0$-plane. The corresponding residue can be obtained from 
Eq.\,(\ref{MIRFphoton1}) by adding an overall minus sign and by substituting
$\beta \to -\beta$ and $D_1 \leftrightarrow D_2$ inside the square brackets:  
\be
  M_{0124}^{\mbox{\scriptsize`$\gamma$'},\,2} = - \bar{M}_{B}\,
               \frac{\alpha}{2\pi}\,\frac{1-\beta x_{2}}{D_{1}}
               \int \frac{d^{2}\Omega_k}{2\pi}\,
               \frac{1}{\eta(x)\,(1-\vec{v}_{2}\!\cdot\!\vec{n}_{k})}\,
               \biggl[ \ln\biggl( \frac{ - 2\beta x + i o}{1+\beta x} \biggr)
                     - \ln(D) + \ln(-D_{2})
               \biggr].
\ee

Next one can perform the integration over the azimuthal angle, with the help 
of the formula
\be
  \int\limits_{0}^{2\pi}\frac{d \phi}{2\pi}\,
  \frac{1}{1-\vec{v}_{i}\!\cdot\!\vec{n}_{k}}
  = 
  \frac{1}{|x - x_{i}|}.
\ee
This expression is a possible source of collinear divergences, which are
regularized by introducing the small non-zero fermion masses. In terms of this
regularization, $|x - x_{i}|$ is replaced by 
$\sqrt{(x - x_{i})^{2} + m_{i}^{2}(1 - x_{i}^{2})/E_{i}^{2}}$. The sum of the 
two photon-pole residues amounts to 
\be
  M_{0124}^{\mbox{\scriptsize`$\gamma$'}} = \bar{M}_{B}\,
               \frac{\alpha}{2\pi}\,\frac{1-\beta x_{2}}{D_{1}}
               \int\limits_{-1}^{1}\frac{d x}{\eta(x)\, |x - x_{2}|}\,
               \biggl\{ \ln \biggl( \frac{1+\beta x}{1-\beta x} \biggr)
                      + \ln \biggl( \frac{D_{1}}{D_{2}} \biggr)
                      + i\pi [\theta(-x) - \theta(x)]
               \biggr\}.
\ee
The last integration gives rise to integrals of logarithmic and dilogarithmic 
type. Let us single out the answer for the $\theta$-function-dependent terms:
\be
  M_{0124}^{\mbox{\scriptsize`$\gamma$'},\,\theta} = \bar{M}_{B}\,
               \frac{\alpha}{2\pi}\,\frac{1-\beta x_{2}}{D_{1}\eta(x_{2})}\,
                i\pi\,\Bigl\{ C_{1} [\theta(-x_{2}) - \theta(x_{2})]
                          + 2 C_{2}\,\theta(x_{2}) + 2 C_{3}\,\theta(-x_{2})
                      \Bigr\},
\ee
where
\begin{eqnarray}
  C_{1} &=& \ln\biggl( \frac{\eta(x_{2})}{\eta(1)} \biggr)
            + \ln\biggl( \frac{\eta(x_{2})}{\eta(-1)} \biggr)
            + \ln\biggl( \frac{4 E_{2}^{2}}{m_{2}^{2}} \biggr), \ \ \ 
  C_{2}  =  \ln\biggl( \frac{D}{\eta(-1)} \biggr) 
            + \ln\biggl( \frac{1+x_{2}}{x_{2}} \biggr), \nonumber \\[1mm]
  C_{3} &=& \ln\biggl( \frac{\eta(1)}{D} \biggr)  
            + \ln\biggl( \frac{x_{2}}{x_{2}-1} \biggr).
\end{eqnarray}
Separately, the particle-pole residue and the photon-pole residues depend on 
the sign of $x_{2}$. However, the sum of these terms does not.
This dependence on $x_{2}$ at the intermediate stage of the calculation is a 
consequence of the decomposition of the unstable $W$-boson propagators. 

The final answer for the contribution of the infrared-finite virtual 
four-point function $D_{0124}$ to the non-factorizable matrix element is given
by
\begin{eqnarray}
\label{MIRFphoton}
  M_{0124} &=& \bar{M}_{B}\,
               \frac{\alpha}{2\pi}\,\frac{1-\beta x_{2}}{D_{1}\eta(x_{2})}\,
               \Biggl\{ \Bigl[ F_{2}(x_{2}|x_{2})-F_{2}(-D_{0}|x_{2}) \Bigr] 
                        \ln\biggl( \frac{D_{1}}{D_{2}} \biggr)
               \Biggr. \nonumber \\[2mm] 
           & & \Biggl.{}- F_{1}\biggl(-D_{0};\beta | x_{2}\biggr)
                        + F_{1}\biggl(-D_{0};-\beta | x_{2}\biggr)
                        + F_{1}\biggl( x_{2};\beta | x_{2}\biggr)
                        - F_{1}\biggl( x_{2};-\beta | x_{2}\biggr)
               \Biggr. \nonumber \\[2mm]  
           & & \Biggl.{}+ i\pi\,\biggl[ 
                           2\ln\biggl( \frac{\eta(x_{2})}{D_{1}} \biggr)
                           + \ln\biggl( \frac{\eta(1)}{\eta(-1)} \biggr)
                           + 2\ln\biggl( \frac{1-\beta x_{2}}
                                              {\beta(1-x_{2})} \biggr) 
                           + \ln\biggl( \frac{E_{2}^{2}}{m_{2}^{2}} \biggr)
                                \biggr]
               \Biggr\},
\end{eqnarray}
where
\be
\label{D0}
  D_{0} = \frac{1}{\beta}\,\frac{D_{1}+D_{2}}{D_{1}-D_{2}}.
\ee
The logarithmic and dilogarithmic functions $F_{1,2}$ can be found in 
App.\,\ref{app:dmi/f1_f2}. The final answer for $M_{0124}$ agrees with the 
answer presented in \cite{melyak}. It is also in complete numerical and 
analytical agreement with the corresponding expression in 
Sect.\,\ref{sec:feynman}, which was calculated with the help of the MST.

The contribution from the other infrared-finite virtual four-point function, 
$D_{0123}$, can be obtained from Eq.\,(\ref{MIRFphoton}) by substituting
$(p_1,k_1) \leftrightarrow (p_2,k_2)$.

\subsubsection{The pure photon-pole part}

As was already explained in Sect.\,\ref{sec:feynman/prescriptions}, the 
photon-pole parts of the virtual scalar functions can be related to the 
corresponding bremsstrahlung interferences. To this end, one needs to 
calculate the pure photon-pole contribution to the matrix element, without
performing the decomposition of the unstable $W$-boson propagators, since 
this decomposition mixes photon- and particle-pole contributions. 

This calculation is pretty much the same as the one discussed in the previous 
subsection. We present only the answer:
\begin{eqnarray}
  M_{0124}^{\gamma} &=& \bar{M}_{B}\,
                \frac{\alpha}{2\pi}\,\frac{1-\beta x_{2}}{D_{1}\eta(x_{2})}\,
                \Biggl\{ \Bigl[ \ln\biggl( \frac{D_{1}}{D_{2}} \biggr) + i\pi
                         \Bigr]
                         \Bigl[ F_{2}(x_{2}|x_{2}) - F_{2}(-D_{0}|x_{2})
                         \Bigr]
                         - F_{1}\biggl(-D_{0};   \beta | x_{2}\biggr)
                \Biggr. \nonumber \\
                    & & 
                \Biggl.{} \hphantom{\bar{M}_{B}\,\frac{\alpha}{2\pi}\,
                                    \frac{1-\beta x_{2}}{D_{1}\eta(x_{2})}\,a}
                         + F_{1}\biggl(-D_{0}; - \beta | x_{2}\biggr)
                         + F_{1}\biggl( x_{2};   \beta | x_{2}\biggr)
                         - F_{1}\biggl( x_{2}; - \beta | x_{2}\biggr)
                \Biggr].  
\end{eqnarray}
Note that the photon pole has been evaluated in the upper half of the complex 
$k_0$-plane. The reason for this lies in the fact that we have opted to perform
the calculations in the most economic way. In this approach 
$D_{0124}^{\gamma}$ is obtained from $D_{0123}^{\gamma}$ by substituting
$(p_1,k_1) \leftrightarrow (p_2,k_2)$, which automatically shifts the 
photon-pole from the lower to the upper half-plane 
(see Sect.\,\ref{sec:feynman/prescriptions}).

\subsection{The infrared-divergent scalar four-point function}
\label{sec:dmi/IRD}

In this subsection we present the calculation of the infrared-divergent 
virtual scalar four-point function $D_{0134}$. In the DMI method such 
functions 
are not needed for the calculation of the non-factorizable corrections. They 
arise only in the form of initial--final state interferences. Such corrections 
vanish when the corresponding bremsstrahlung interferences are taken into 
account, as was explained in Sect.\,\ref{sec:feynman/prescriptions}. We perform
the calculation mainly to study how one can handle infrared and collinear 
divergences in the DMI scheme and to provide an independent check of the 
results obtained in Sect.\,\ref{sec:feynman}.

The infrared-divergent virtual scalar four-point function $D_{0134}$ is 
defined as
\be
  D_{0134} = \int\frac{d^{4}k}{(2\pi)^{4}}\,
             \frac{1}{[k^{2}-\lambda^{2}+io] [D_{1}-2(p_{1}\cdot k)] 
                       [ - 2(k_{1}\cdot k) + io] [2(k_{2}\cdot k) + i o]}.
\ee
We regularize the infrared divergences by introducing a regulator mass 
$\lambda$ for the photon. There are also collinear divergences, which are 
regularized by the small non-zero fermion masses.

The pole structure of this integral is such that no propagator decomposition 
is required. There is one photon pole in each of the half-planes of the 
complex variable $k_0$. There are two particle poles in the upper half-plane, 
and only one in the lower half-plane. Therefore, we opt to close the 
integration contour in the lower half-plane, resulting in only two 
contributions to the scalar function: one photon-pole residue and one 
particle-pole residue.

\subsubsection{Particle-pole contribution}

One can proceed in the same way as in Sect.\,\ref{sec:dmi/IRF}. We take the
residue at the particle pole $k_0 = \vec{v}_{2}\!\cdot\!\vec{k}$ and
exponentiate the propagators by introducing an integration over ``time'':
\be
  D^{\mbox{\scriptsize part}}_{0134}  = \frac{i}{8 E E_{1} E_{2}}
               \int\limits_{0}^{\infty} d\tau\, d t\,
               e^{i \tau \frac{D_{1}}{2 E}}
               \int \frac{d^{3}k}{(2\pi)^{3}}\,
               \frac{e^{i \vec{r}\cdot\vec{k}}}
                    {(\vec{k}\!\cdot\!\vec{v}_{2})^2 - \vec{k}^2 - \lambda^2},
\ee
where 
$\vec{r} = \tau\,(\vec{v}-\vec{v}_{2}) + t\,(\vec{v}_{1}-\vec{v}_{2})$. 
Again we can perform the integration over the momentum $\vec{k}$, which is
similar to the $\lambda$-screened relativistic Coulomb potential 
of a moving particle. As the scalar function
is infrared-divergent, one should keep the photon mass $\lambda$. The result 
of the integration is
\be
\label{philambda}
  \phi_{\lambda}(r) = -4\pi\int \frac{d^{3}k}{(2\pi)^{3}}\,
                      \frac{e^{i \vec{r}\cdot\vec{k}}}
                           {(\vec{k}\!\cdot\!\vec{v}_{2})^2 - \vec{k}^2
                            - \lambda^2}
                    = \frac{ e^{-\,\frac{\lambda}{\sqrt{1-v_{2}^{2}}}
                        \sqrt{r_{\parallel}^{2} + r_{\perp}^{2}(1-v_{2}^{2})}}}
                      {\sqrt{r_{\parallel}^{2} + r_{\perp}^{2}(1-v_{2}^{2})}}.
\ee
Here both $\lambda$ and $1-v_{i}^{2}$ are small. We will consider the limit 
$\lambda\to 0$ and $v_{i}\to 1$, such that $\lambda \ll \sqrt{1-v_{i}^{2}}$.  

The particle-pole contribution now takes the form
\be
    D^{\mbox{\scriptsize part}}_{0134}  = - \frac{i}{32 \pi E E_{1} E_{2}}
               \int\limits_{0}^{\infty} d\tau\, d t\,
               e^{i \tau \frac{D_{1}}{2 E}}\,    
               \frac{e^{-\,\frac{\lambda}{\sqrt{1-v_{2}^{2}}}
                     \sqrt{r_{\parallel}^{2} + r_{\perp}^{2}(1-v_{2}^{2})}}}
                    {\sqrt{r_{\parallel}^{2} + r_{\perp}^{2}(1-v_{2}^{2})}},
\ee
where
$r_{\parallel}^{2} + r_{\perp}^{2}(1-v_{2}^{2}) = a + b t + c t^{2}$,
with coefficients
\begin{eqnarray}
  a &=& \tau^{2}(1-\beta x_{2})^{2} 
        + \frac{m_{2}^{2}}{E_{2}^{2}}\, \tau^{2}\beta^{2}\sin^{2}\theta_{2}, 
        \nonumber \\
  b &=& 2\tau (1-x_{12}) (1 - \beta x_{2}) 
        + 2\,\frac{m_2^2}{E_2^2}\, \tau\beta \sin\theta_{2}\sin\theta_{12},
        \nonumber \\
  c &=& (1-x_{12})^{2} + \frac{m_{2}^{2}}{E_{2}^{2}}\sin^{2}\theta_{12}.
\end{eqnarray}
The integral over $t$ is logarithmically divergent in $\lambda$:
\be
  I_{\lambda} = \int\limits_{0}^{\infty}d t \,     
                \frac{ e^{-\,\frac{\lambda}{\sqrt{1-v_{2}^{2}}}
                   \sqrt{a + b t + c t^{2}}}}{\sqrt{a + b t + c t^{2}}}
        \approx \frac{1}{\sqrt{c}}\,\biggl[ -{{\bf C}} 
                         + \ln\biggl( \frac{4\sqrt{c}\,m_2}
                           {\lambda E_2\,(b + 2\sqrt{a c}\,)} \biggr)
                                    \biggr],
\ee
where ${{\bf C}}$ is the Euler constant.

The last integration over $\tau$ is relatively simple, yielding
\be
  D_{0134}^{\mbox{\scriptsize part}} = 
               \frac{1}{8\pi s_{12}}\,\frac{1}{D_1}
               \Biggl[ \ln\biggl( \frac{D_1}{i M_W^2} \biggr)
                       - \ln\biggl( \frac{s_{211^{\prime}}}{M_W^2} - 1 \biggr)
                       - \ln\biggl( \frac{\lambda}{m_2} \biggr)
               \Biggr].
\ee
The invariants $s_{12}$ and $s_{211^{\prime}}$ are defined in 
Eq.\,(\ref{invariants}).

\subsubsection{Photon-pole contribution}

Next the photon-pole residue at $k_0=\omega=\sqrt{\vec{k}^2+\lambda^2-io}\,$
is determined:
\be
  D_{0134}^{\gamma} = -\frac{i}{16 E E_{1} E_{2}}
           \int \frac{d^{3}k}{(2\pi)^{3}\omega}\,
           \frac{1}{(1-\beta x)[\omega - \frac{D_{1}}{2 E (1 - \beta x)}] 
                    [\omega - |\vec{k}|\,(\vec{n}_k\!\cdot\!\vec{v}_{1})] 
                    [\omega - |\vec{k}|\,(\vec{n}_k\!\cdot\!\vec{v}_{2})]}.
\ee
We want to keep the propagators 
$[\omega - |\vec{k}|\,(\vec{n}_k\!\cdot\!\vec{v}_{i})]$ as they are, instead 
of writing them as $|\vec{k}|\,[1-\vec{n}_k\!\cdot\!\vec{v}_{i}]$, as was 
done in Ref.\,\cite{melyak}. Keeping the exact form of the propagators leads
to double-logarithmic collinear divergences. If, instead, the simplified 
version were to be used, then the double-logarithmic terms would be lost, and 
one cannot be sure whether the single-logarithmic divergence and the finite 
part would be correct.

First we perform the integration over $|\vec{k}|$. The presence of $\lambda$
in the light-fermion propagators complicates things considerably. The 
light-fermion propagators can be rewritten in the following way:
\be
\label{fermionsplitup}
  \frac{1}{[\omega - |\vec{k}|\,(\vec{n}_k\!\cdot\!\vec{v}_{1})]
           [\omega - |\vec{k}|\,(\vec{n}_k\!\cdot\!\vec{v}_{2})]}
  = 
  \frac{1}{|\vec{k}|\,(\vec{n}_k\!\cdot\!\vec{v}_{1}
                       - \vec{n}_k\!\cdot\!\vec{v}_{2})}\,
  \biggl[\, \frac{1}{ \omega - |\vec{k}|\,(\vec{n}_k\!\cdot\!\vec{v}_{1})}
          - \frac{1}{ \omega - |\vec{k}|\,(\vec{n}_k\!\cdot\!\vec{v}_{2})}\,
  \biggr].
\ee
After the integration over $|\vec{k}|$ the photon-pole contribution will
be of the form
\be
  D_{0134}^{\gamma} = \frac{i}{16\pi^{2} E_{1} E_{2} D_{1}}\,
                      \bigl[ I_{0} + I_{1} + I_{2} \bigr],
\ee
where
\begin{eqnarray}
  I_{0} &=& \int \frac{d^{2}\Omega_k}{4\pi}\,
            \frac{1}{[1 - \vec{n}_k\!\cdot\!\vec{v}_{1}]
                     [1 - \vec{n}_k\!\cdot\!\vec{v}_{2}]}\,
            \ln\biggl( \frac{-D_{1}}{E\lambda (1 - \beta x)} \biggr),
            \nonumber \\[1mm]
  I_{1} &=& {\cal P}\,\Biggl\{ \int \frac{d^{2}\Omega_k}{4\pi}\,
            \frac{1}{[\vec{n}_k\!\cdot\!\vec{v}_{1} 
                      - \vec{n}_k\!\cdot\!\vec{v}_{2}]
                     [1 - (\vec{n}_k\!\cdot\!\vec{v}_{1})^2]}
            \ln\biggl( \frac{1 - \vec{n}_k\!\cdot\!\vec{v}_{1}}{2} \biggr)
                      \Biggr\}.
\end{eqnarray}
The expression for $I_{2}$ can be obtained from $I_{1}$ by the substitution 
$\vec{v}_1 \leftrightarrow \vec{v}_2$. The integral $I_{0}$ is similar to the
one that shows up in the approach of \cite{melyak}. It contains only  
single-logarithmic collinear divergences. The integrals $I_{1,2}$ are new and 
give rise to double-logarithmic terms. They are evaluated as a principal-value
integral, since the singularity present in 
$1/[\vec{n}_k\!\cdot\!\vec{v}_{1} - \vec{n}_k\!\cdot\!\vec{v}_{2}]$ is an 
artefact of the split-up (\ref{fermionsplitup}) and disappears in the sum 
$I_1+I_2$. 

As a next step we integrate over the azimuthal angle $\phi$. We obtain for the 
first integral
\be
  I_{0} = \frac{1}{2 (1 - x_{12}^{2})}\int\limits_{-1}^{+1}
          \frac{d x}{(x - x_{a})(x - x_{b})}\,
          \biggl[ \frac{J_{1} - x K_{1}}{|x - x_{1}|}
                  + \frac{J_{2} - x K_{2}}{|x - x_{2}|}               
          \biggr]\,
          \ln\biggl( \frac{-D_{1}}{E\lambda (1 - \beta x)} \biggr),
\ee
with
\begin{eqnarray}
  x_{a,b} &=& \frac{x_{1}+x_{2}\pm i\sin\theta_{1}\sin\theta_{2}\sin\phi_{12}}
                   {1 + x_{12}}, \nonumber \\[1mm]
  J_{1}   &=& 1-x_{12} - x_{1}(x_{1} - x_{2}), \ \ \
              K_{1} = x_{2} - x_{1} x_{12}, \nonumber \\[1mm]   
  J_{2}   &=& 1-x_{12} - x_{2}(x_{2} - x_{1}), \ \ \ 
              K_{2} = x_{1} - x_{2} x_{12}.
\end{eqnarray}
As indicated in Sect.\,\ref{sec:dmi/IRF}, $|x - x_{i}|$ should be regularized 
by keeping the small non-zero fermion masses:
$|x - x_{i}| \to \sqrt{(x - x_{i})^{2} + \mu_i^{2}}$, with
$\mu_i^{2} = m_{i}^{2}(1 - x_{i}^{2})/E_{i}^{2}$. Using the result for the 
principal-value integral given in App.\,\ref{app:dmi/azimuthal}, we find for
the second integral
\be
  I_{1} = \frac{1}{4 |\vec{v}_{1} - \vec{v}_{2}|}
          \int\limits_{x_{+}}^{1} \frac{d x}{\sqrt{x^{2} - x_{+}^{2}}}\,
          \biggl[ \frac{1}{1-v_{1} x} + \frac{1}{1+v_{1} x} \biggr]
          \ln\biggl( \frac{1-v_{1} x}{1+v_{1} x} \biggr),
\ee
where $x_{\pm}=\sqrt{(1\pm x_{12})/2}\,$.

One is left with a one-dimensional integration over $x$. The integrals 
$I_{1,2}$ can be expressed in terms of the dilogarithmic functions $\ff_1$ and
$\ff_2$, defined in App.\,\ref{app:dmi/ff1_ff2}:
\be
  I_{1,2} = \frac{1}{8 x_{-}}\,\bigl[\ff_{1}(x_{+};v_{1,2}) 
                                   + \ff_{2}(x_{+})\bigr].
\ee
The answer for the integral $I_{0}$ is
\begin{eqnarray}
  I_0 &=& \frac{1}{1-x_{12}}\,\ln\biggl( \frac{E\lambda \beta}{-D_{1}} \biggr)
                              \ln\biggl( \frac{m_{1}m_{2}}{s_{12}} \biggr)
          - \sum_{i=1}^2 \frac{x_{i} - x_{a}}{4(1-x_{12})}\,
            {\cal K}\bigl(-x_{a};\frac{1}{\beta} | x_{i};\mu_{i}^{2}\bigr)
          \nonumber \\[1mm]
      & & {}- \sum_{i=1}^2 \frac{x_{i} - x_{b}}{4(1-x_{12})}\,
            {\cal K}\bigl(-x_{b};\frac{1}{\beta} | x_{i};\mu_{i}^{2}\bigr),
\end{eqnarray}
where the function ${\cal K}$ is introduced in App.\,\ref{app:dmi/ff1_ff2}.

The complete expression for the infrared-divergent scalar four-point function 
is obtained as the sum of the particle-pole and photon-pole contributions. 
When comparing all the above results with the ones obtained in 
Sects.\,\ref{sec:feynman/IRD} and \ref{sec:feynman/IRDreal}, complete 
numerical agreement is found for both virtual and real four-point functions,
including collinear and infrared divergences.

\subsection{Non-factorizable corrections from the five-point functions}

In this subsection we describe the calculation of the scalar five-point 
functions using the DMI method. The contribution of the virtual five-point 
function to the non-factorizable matrix element is given by
\be
  M = i\bar{M}_{B}\,\int\frac{d^{4}k}{(2\pi)^{4}}\,
      \frac{16\pi\alpha\,(k_{1}\cdot k_{2})}{[k^{2} - \lambda^{2}+io]
            [-2(k\cdot k_{1})+io] [2(k\cdot k_{2})+io]
            [D_{1} - 2(k\cdot p_{1})][D_{2} + 2(k\cdot p_{2})]}.
\ee
In analogy to Sect.\,\ref{sec:dmi/IRF}, one can perform a decomposition 
of the unstable $W$-boson propagators  
\be
\label{Wdecomp2}
  \frac{1}{[D_{1}-2(p_{1}\cdot k)][D_{2}+2(p_{2}\cdot k)]}
  =
  \biggl[ \frac{1}{D_{1}-2(p_{1}\cdot k)} + \frac{1}{D_{2}+2(p_{2}\cdot k)}
  \biggr] \frac{1}{D+4\pvec\!\cdot\!\vec{k}}\, .
\ee
In this way the matrix element splits into two terms. If the
integration contour in the complex $k_0$-plane is chosen properly, each term
involves one photon-pole contribution and one particle-pole contribution.

\subsubsection{Particle-pole residues}

We first calculate the particle-pole residue that contributes to the first 
term in Eq.\,(\ref{Wdecomp2}). We proceed in the usual way, by taking the 
residue at $k_0 = \vec{v}_2\!\cdot\!\vec{k}$ and subsequently  
exponentiating the propagators. In this case this procedure requires three
integrations:
\be
  M^{\mbox{\scriptsize `part'}}_2 = \bar{M}_{B}\,\frac{i\pi \alpha}{2}\,
             \frac{1-x_{12}}{E^{2}}\,\int\limits_0^\infty d t\,d\tau\,d\tau_1\,
             e^{i\bigl[\frac{D}{4E}\tau + \frac{D_{1}}{2 E}\tau_{1}\bigr]}
             \int\frac{d^{3}k}{(2\pi)^{3}}\,
             \frac{e^{i \vec{r}_2\cdot\vec{k}}}
                    {(\vec{k}\!\cdot\!\vec{v}_{2})^2 - \vec{k}^2 - \lambda^2},
\ee
where $\vec{r}_2 = \tau\,\vec{v} + \tau_1\,(\vec{v}-\vec{v}_{2}) 
                   + t\,(\vec{v}_{1}-\vec{v}_{2})$. 
The integral over $d^{3}k$ 
is the same as the one evaluated for the calculation of the infrared-divergent 
four-point function [see Eq.\,(\ref{philambda})]. Again, a non-zero photon 
mass $\lambda$ is needed for the regularization of the infrared divergences.
The particle-pole residue now amounts to
\be
  M^{\mbox{\scriptsize `part'}}_2 = -\bar{M}_{B}\,\frac{i \alpha}{8}\,
            \frac{1-x_{12}}{E^{2}}\,\int\limits_0^\infty d t\,d\tau\,d\tau_1\,
            e^{i\bigl[\frac{D}{4E}\tau + \frac{D_{1}}{2 E}\tau_{1}\bigr]}\,
            \frac{e^{-\,\frac{\lambda}{\sqrt{1-v_{2}^{2}}}
                     \sqrt{r_{\parallel}^{2} + r_{\perp}^{2}(1-v_{2}^{2})}}}
                 {\sqrt{r_{\parallel}^{2} + r_{\perp}^{2}(1-v_{2}^{2})}},
\ee
where $r_{\parallel}^{2} + r_{\perp}^{2}(1-v_{2}^{2}) = a + b t + c t^{2}$,
with coefficients
\begin{eqnarray}
  a &=& \bigl[\tau \beta x_{2} - \tau_{1}(1-\beta x_{2})\bigr]^{2}
        + \frac{m_{2}^{2}}{E_{2}^{2}}\, (\tau+\tau_{1})^{2}\beta^{2}
          \sin^{2}\theta_{2}, \nonumber \\
  b &=& - 2(1-x_{12}) \bigl[\tau\beta x_{2} - \tau_{1}(1-\beta x_{2})\bigr]
        + 2\,\frac{m_{2}^{2}}{E_{2}^{2}}\,\beta(\tau+\tau_{1})
          \sin\theta_{12}\sin\theta_{2}, \nonumber \\
  c &=& (1-x_{12})^{2} + \frac{m_{2}^{2}}{E_{2}^{2}}\sin^{2}\theta_{12}.
\end{eqnarray}
Following Sect.\,\ref{sec:dmi/IRD}, first the integration over $t$ is 
performed, yielding the logarithmically-divergent result
\be
  M^{\mbox{\scriptsize `part'}}_2 = -\bar{M}_{B}\, \frac{i \alpha}{8 E^{2}}
            \int\limits_0^\infty d \tau\, d \tau_{1}\,
            e^{i\bigl[\frac{D}{4E}\tau + \frac{D_{1}}{2 E}\tau_{1}\bigr]}
            \Biggl[ -{{\bf C}} 
                    + \ln\biggl( \frac{4\sqrt{c}\,m_2}
                      {\lambda E_{2}\,(b+2\sqrt{a c}\,)} \biggr)
            \Biggr],
\ee
To linearize $\, b+2\sqrt{a c} \,$ with respect to one of the integration
variables, one should make a change of variables according to
$(\tau,\tau_{1})\to (\xi,y)$, with $\tau = \xi y$, $\tau_{1} = \xi (1 - y)$.
In this way, the integration over $\xi$ can be trivially performed:
\be
  M^{\mbox{\scriptsize `part'}}_2\!\!\!= \bar{M}_{B}\,\frac{i\alpha}{8 E^{2}}\,
          \int\limits_{0}^{1}
          \frac{d y}{\bigl[\frac{D}{4E}y + \frac{D_{1}}{2E}(1-y)\bigr]^{2}}\,
          \Biggl\{ -1 + \ln\biggl( \frac{4m_{2}(1-x_{12})}
                   {\lambda E_{2}\,(b^{\prime}+2\sqrt{a^{\prime}c^{\prime}})}
                           \biggr)
                   + \ln\biggl( \frac{Dy+2D_1(1-y)}{4iE} \biggr)
          \Biggr\},
\ee
where the coefficients $a^{\prime},b^{\prime}$, and $c^{\prime}$ follow from 
the coefficients $a,b$, and $c$, by substituting $\tau\to y$ and 
$\tau_{1}\to (1-y)$.

The last integration is technically quite involved, but only gives rise to
logarithms. Note that one should carefully analyse
the infrared and collinear 
divergences, present in this integral. The final answer is formally different 
for $x_{2}<0$ and $x_{2}>0$. This is the same phenomenon as observed in 
Sect.\,\ref{sec:dmi/IRF}, which can be attributed to the decomposition of the 
$W$-boson propagators. The result for the photon-pole residue will compensate
this dual behaviour, leading to a combined result that is analytically the 
same for both $x_{2}<0$ and $x_{2}>0$. In a similar way the second 
particle-pole residue, corresponding to the second term in 
Eq.\,(\ref{Wdecomp2}), will formally depend on the sign of $x_1$.

We will not bore the reader with all the different cases and merely present the
answer for the case $x_1>0,\,x_{2}<0$. Taking into account both particle-pole 
residues, the final answer reads 
\begin{eqnarray}
\label{M01234part}
  \underline{x_1>0,\,x_2<0}: \hspace*{-1.9cm}&& \nonumber \\[1mm]
  M^{\mbox{\scriptsize `part'}} \!\!&=&\!\! {}-\bar{M}_{B}\,\frac{i\alpha}{2}\,
     \Biggl\{ \frac{2}{D_{1}D}\ln\biggl( \frac{2\lambda E}{M_W^2} \biggr)
            + \frac{2}{D_{1}D}\ln\biggl( \frac{E_{2}}{m_{2}} \biggr)
            + \frac{2}{D_{2}D}\ln\biggl( \frac{D}{i M_{W}^{2}} \biggr)
            - \frac{2}{D_1 D_2}\ln\biggl( \frac{D_{1}}{iM_W^2} \biggr)
     \Biggr. \nonumber \\[1mm]
                                \!\!& &\!\!
     \Biggl.\hphantom{- \bar{M}_{B}\,\frac{i \alpha}{2}\,A}
          {}+ \frac{2(1-\beta x_{2})}{D_{1}\eta(x_{2})}\ln(1-\beta x_{2})
            - \frac{2(1+\beta x_{2})}{D_{2}\eta(x_{2})}
              \ln\biggl( \frac{D}{D_{1}} \biggr)
            + \frac{4\beta x_{2}}{D\eta(x_{2})}\ln(-2\beta x_{2})
     \Biggr\} \nonumber \\[1mm]
                                \!\!& &\!\! {}-\bar{M}_{B}\,\frac{i\alpha}{2}\,
     \Biggl\{ x_2\to x_1,\ m_2\to m_1,\ E_2\to E_1,\
              D_1 \leftrightarrow D_2,\ \beta \to -\beta 
     \Biggr\}.
\end{eqnarray}
Note that the terms $\ln\bigl(\frac{E_{2}}{m_{2}}\bigr)$ and
 $\ln\bigl(\frac{E_{1}}{m_{1}}\bigr)$ cause the difference with the 
results presented in Ref.\,\cite{melyak}.

\subsubsection{Photon-pole residues} 

Next we determine the photon-pole residues. Each of the terms in the propagator
decomposition (\ref{Wdecomp2}) gives rise to one photon-pole residue, situated 
at $k_0 = \omega = \pm\sqrt{\vec{k}^2+\lambda^2-io}$. In the same way as in
Sect.\,\ref{sec:dmi/IRD}, the light-fermion propagators occurring in the 
photon-pole residues can be rewritten according to Eq.\,(\ref{fermionsplitup}).
Again we introduce spherical coordinates, with the polar axis defined along 
$\pvec$. For the integration over $|\vec{k}|$ we keep the $\lambda$ dependence
of $\omega$ in order to get the correct divergences. The combined result of all
photon-pole residues is given by 
\be
\label{eq:dmi/5p/1}
  M^{\mbox{\scriptsize`$\gamma$'}} = - \bar{M}_{B}\,\frac{\alpha}{\pi}\,
         (1 - x_{12}) \int\frac{d^{2}\Omega_k}{4\pi}
         \Biggl\{ \frac{\Psi(D_{1},D_{2},x)}{(1-\alpha_{1})(1 - \alpha_{2})}
                + \frac{1}{D_{1}D_{2}} \Bigl[ \Phi(\alpha_{1},\alpha_{2})
                                            + \Phi(\alpha_{2},\alpha_{1})\Bigr]
         \Biggr\},
\ee
with
\begin{eqnarray}
  \Psi(D_{1},D_{2},x) &=& \Psi_{0} +\Psi_{12} +\Psi_{\theta}, \nonumber \\[1mm]
  \Psi_{0} &=& - \frac{1}{D_{1}D_{2}}\,
               \Bigl[ \ln\biggl( \frac{\lambda E}{M_W^2} \biggr) + i\pi \Bigr],
               \nonumber \\[1mm]
  \Psi_{12}&=& \frac{1-\beta x}{D_{1}\eta(x)}\,
               \ln\biggl( \frac{D_{1}}{M_W^2\,(1-\beta x)} \biggr)
             + \frac{1+\beta x}{D_{2}\eta(x)}\,
               \ln\biggl( \frac{D_{2}}{M_W^2\,(1+\beta x)} \biggr), 
               \nonumber \\[1mm]
  \Psi_{\theta} &=&  \frac{2\beta x}{D\eta(x)}\, i\pi\,
               \bigl[ \theta(\beta x)- \theta(-\beta x) \bigr], 
               \nonumber \\[1mm]
  \Phi(\alpha_{1},\alpha_{2}) &=& 
               \frac{1}{(\alpha_{1}-\alpha_{2})(1-\alpha_{1}^{2})}
               \ln\biggl( \frac{1-\alpha_{1}}{2} \biggr),
\end{eqnarray}
and $\alpha_{i}=(\vec{n}_k\!\cdot\!\vec{v}_{i})$ for $(i=1,2)$.

The $\Psi$-term in Eq.\,(\ref{eq:dmi/5p/1}) also emerges in the calculations 
presented in \cite{melyak}, up to the divergent term $\ln(\lambda E)$. This 
$\Psi$-term contains infrared divergences and logarithmic collinear 
divergences.
The other two terms in Eq.\,(\ref{eq:dmi/5p/1}) are of a type that was already
encountered in Sect.\,\ref{sec:dmi/IRD}. They will give rise to  
double-logarithmic collinear divergences

As in Sect.\,\ref{sec:dmi/IRD}, we proceed by performing the azimuthal 
integration. The remaining integration over $x=\cos\theta$ gives logarithms 
and dilogarithms. Most of the ingredients of this final step in the calculation
have already been discussed in the previous subsections. Therefore we only
give the answer. First we do so for the $\Psi_{\theta}$-terms.
As was observed for the particle-pole residues, the results depend on the sign
of $x_{1,2}$. Adopting the same sign choice as in Eq.\,(\ref{M01234part}), we
obtain
\begin{eqnarray}
\label{eq:dmi/5p/Mgtheta}
  \!\!\!\!\underline{x_1>0,\,x_2<0}: \hspace*{-2cm}&& \nonumber \\[1mm]
  M^{\mbox{\scriptsize`$\gamma$'}}_{\theta} &=& 
             \bar{M}_{B}\, \frac{\alpha}{2\pi}\, i\pi\, \frac{1}{D}\,
             \Biggl\{ \R_{\eta,2}\biggl[ 
                          \ln\biggl( \frac{\eta(-1)}{\eta(1)} \biggr)
                          - 2\ln\biggl( \frac{\eta(x_{2})}{D} \biggr)
                                 \biggr]
                      + \R_{2}\biggl[
                          \ln\biggl( \frac{4E_{2}^{2}}{m_{2}^{2}} \biggr) 
                          + 2\ln\biggl( \frac{-x_{2}}{1-x_{2}} \biggr)
                              \biggr]
             \Biggr.  \nonumber \\
                                            & & 
             \Biggl.\hphantom{\bar{M}_{B}\, \frac{\alpha}{2\pi}\, i\pi\, 
                              \frac{1}{D}\,a}
                    {}+ \sum_{j=a,b} \R_{j}\biggl[ 
                          \ln\biggl( \frac{-1-x_{j}}{1 - x_{j}} \biggr)
                          - 2\ln\biggl( \frac{x_{2} - x_{j}}{-x_{j}} \biggr)
                                           \biggr]
             \Biggr\} \nonumber \\
                                            &\,\,\,+\!\!\!& 
             \bar{M}_{B}\, \frac{\alpha}{2\pi}\, i\pi\, \frac{1}{D}\,
             \Biggl\{ \R_{\eta,1}\biggl[
                          \ln\biggl( \frac{\eta(-1)}{\eta(1)} \biggr)
                          + 2\ln\biggl( \frac{\eta(x_{1})}{D} \biggr)
                                 \biggr]
                      - \R_{1}\biggl[ 
                          \ln\biggl( \frac{4E_{1}^{2}}{m_{1}^{2}} \biggr)
                          + 2\ln\biggl( \frac{x_{1}}{1+x_{1}} \biggr)
                              \biggr] 
             \Biggr.  \nonumber \\
                                            & &
             \Biggl.\hphantom{\bar{M}_{B}\, \frac{\alpha}{2\pi}\, i\pi\, 
                              \frac{1}{D}\,a}
                    {}+ \sum_{j=a,b} \R_{j}\biggl[
                          \ln\biggl( \frac{-1-x_{j}}{1 - x_{j}} \biggr)
                          + 2\ln\biggl( \frac{x_{1} - x_{j}}{-x_{j}} \biggr)
                                           \biggr]
          \Biggr\}.
\end{eqnarray}
The coefficients $\R$ are given by ($i=1,2$) 
\be
  \R_{\eta,i} = 2 \beta D\,\frac{K_{i}\,D + \beta J_{i}\,(D_{1}-D_{2})}
                           {(1+x_{12})\,\eta(x_{a})\,\eta(x_{b})\,\eta(x_{i})},
  \ \ \ 
  \R_{i} = \frac{2\beta x_{i}}{\eta(x_{i})}
  \ \ \ \mbox{and}\ \ \ 
  \R_{a,b} = - \frac{\beta x_{a,b}}{\eta(x_{a,b})}.
\ee
As in Sect.\,\ref{sec:dmi/IRF}, only the sum of the particle-pole residues 
and $M^{\mbox{\scriptsize`$\gamma$'}}_{\theta}$ is independent of the sign of
$x_{1,2}$. 

The evaluation of the remaining terms in Eq.\,(\ref{eq:dmi/5p/1}) is 
straightforward. The answers for the $\Psi_{0}\,$- and $\Phi$-terms are given 
by
\begin{eqnarray}
\label{eq:dmi/5p/Mg0}
  M^{\mbox{\scriptsize`$\gamma$'}}_0 &=&
        -\bar{M}_{B}\, \frac{\alpha}{\pi}\, \frac{1}{D_{1}D_{2}}\,
        \ln\biggl( \frac{m_{1}m_{2}}{s_{12}} \biggr)\,
        \biggl[ \ln\biggl( \frac{\lambda E}{M_{W}^{2}} \biggr) + i\pi \biggr],
        \\[1mm]
\label{eq:dmi/5p/Mf4}
  M^{\mbox{\scriptsize`$\gamma$'}}_{\phi} &=& 
         -\bar{M}_{B}\, \frac{\alpha}{2\pi}\, \frac{1-x_{12}}{D_{1}D_{2}}\,
         \frac{1}{4 x_{-}}\,\bigl[ \ff_{1}(x_{+};v_{2}) + \ff_{1}(x_{+};v_{1})
                                   + 2\,\ff_{2}(x_{+})
                            \bigr].
\end{eqnarray}
The functions $\ff_{1}$ and $\ff_{2}$ can be found in 
App.\,\ref{app:dmi/ff1_ff2}.

Finally, the answer for the $\Psi_{12}\,$-term reads
\be
    M^{\mbox{\scriptsize`$\gamma$'}}_{12} = M^{\mbox{\scriptsize`$\gamma$'}}_1
                     +   M^{\mbox{\scriptsize`$\gamma$'}}_2,
\ee
with
\begin{eqnarray}
\label{eq:dmi/5p/M24}
  M^{\mbox{\scriptsize`$\gamma$'}}_2 &=&
         {}-\bar{M}_{B}\, \frac{\alpha}{2\pi}\, 
         \Biggl\{ \frac{1-\beta x_{2}}{D_{1}\eta(x_{2})}\,
                  \biggl[ \ln\biggl( \frac{4E_{2}^{2}}{m_{2}^{2}} \biggr)
                          \ln\biggl( \frac{D_{1}}{M_{W}^{2}} \biggr)
                          - F_{1}(x_{2};-\beta|x_{2})
                  \biggr]
         \Biggr. \nonumber \\[1mm]
                                     & &
         \Biggl. \hphantom{-\bar{M}_{B}\, \frac{\alpha}{2\pi}\,A}
              {}- \sum\limits_{i=a,b}\frac{1-\beta x_{i}}{2 D_{1}\eta(x_{i})}\,
                  \biggl[ \ln\biggl( \frac{D_{1}}{M_{W}^{2}} \biggr)\,
                          F_{2}(x_{i}|x_{2}) - F_{1}(x_{i};-\beta|x_{2})
                  \biggr] 
         \Biggr. \nonumber \\[1mm]
                                     & &
         \Biggl. \hphantom{-\bar{M}_{B}\, \frac{\alpha}{2\pi}\,A}
              {}- \frac{R_{\eta,2}}{D}\,
                  \biggl[ \ln\biggl( \frac{D_{1}}{M_{W}^{2}} \biggr)\,
                          F_{2}(-D_{0}|x_{2}) - F_{1}(-D_{0};-\beta|x_{2})
                  \biggr]
         \Biggr\} \nonumber \\[1mm]
                                     & &
         {}-\bar{M}_{B}\, \frac{\alpha}{2\pi}\, 
         \Biggl\{ D_{1}\leftrightarrow D_{2},\ \beta\to -\beta
         \Biggr\}.
\end{eqnarray}
Here $D_{0}$ is defined in Eq.\,(\ref{D0}) and the functions $F_{1}$ and 
$F_{2}$ are given in App.\,\ref{app:dmi/f1_f2}. 
Note that the coefficient $R_{\eta,2}$ depends on $\beta$.
The contribution 
$M^{\mbox{\scriptsize`$\gamma$'}}_1$ can be obtained by substituting 
$(E_2,\,m_2,\,x_2) \leftrightarrow (E_1,\,m_1,\,x_1)$ in 
Eq.\,(\ref{eq:dmi/5p/M24}). 

The final answer for the contribution of the virtual five-point function to 
the non-factorizable matrix element can be obtained as
\be
  M = M^{\mbox{\scriptsize `part'}} + M^{\mbox{\scriptsize`$\gamma$'}}_{\theta}
      + M^{\mbox{\scriptsize`$\gamma$'}}_0 
      + M^{\mbox{\scriptsize`$\gamma$'}}_{12}
      + M^{\mbox{\scriptsize`$\gamma$'}}_{\phi},
\ee
with the various contributions given by Eqs.\,(\ref{M01234part}) and  
(\ref{eq:dmi/5p/Mgtheta})--(\ref{eq:dmi/5p/M24}).  
This answer was compared numerically with the corresponding MST-expression in
Sect.\,\ref{sec:decomp}, which was derived by means of a decomposition of the 
five-point function into a sum of four-point functions.
A complete numerical agreement was observed.

\subsubsection{Pure photon-pole part}

In order to calculate the real-photon radiative interference corresponding to 
the five-point function, one has to determine the photon-pole residue in the 
lower half-plane, without performing the propagator decomposition.
The calculation is more or less the same as the one discussed in the previous 
subsection. 

The answer can be written as 
\be
  M^{\gamma} = M^{\gamma}_0 + M^{\gamma}_1 + M^{\gamma}_2 
               + M^{\mbox{\scriptsize`$\gamma$'}}_{\phi}.
\ee
Note that the $M^{\mbox{\scriptsize`$\gamma$'}}_{\phi}$ contribution is the 
same as before [see Eq.\,(\ref{eq:dmi/5p/Mf4})]. The other contributions are
changed slightly:
\be
  M^{\gamma}_0 = -\bar{M}_{B}\, \frac{\alpha}{\pi}\, \frac{1}{D_{1}D_{2}}\,
                 \ln\biggl( \frac{m_{1}m_{2}}{s_{12}} \biggr)\,
                 \ln\biggl( \frac{\lambda E}{M_{W}^{2}} \biggr),
\ee
and 
\begin{eqnarray}
  M^{\gamma}_2 &=& M^{\mbox{\scriptsize`$\gamma$'}}_2  
        + \bar{M}_{B}\, \frac{\alpha}{2\pi}\,i\pi\, 
        \Biggl\{ \frac{1-\beta x_{2}}{D_{1}\eta(x_{2})}\,
                 \ln\biggl( \frac{4E_{2}^{2}}{m_{2}^{2}} \biggr)
               - \sum\limits_{i=a,b}\frac{1-\beta x_{i}}{2D_{1}\eta(x_{i})}\,
                 F_{2}(x_{i}|x_{2})
        \Biggr. \nonumber \\[1mm]
               & &
        \Biggl. \hphantom{M^{\mbox{\scriptsize`$\gamma$'}}_2  
                          + \bar{M}_{B}\, \frac{\alpha}{2\pi}\,i\pi\,A}
             {}- \frac{R_{\eta,2}}{D}\,F_{2}(-D_{0}|x_{2})
        \Biggr\}.
\end{eqnarray}
The contribution $M^{\gamma}_1$ can be obtained by 
substituting $(E_2,\,m_2,\,x_2) \leftrightarrow (E_1,\,m_1,\,x_1)$.

\section{Complete results}
\label{sec:results}

Up to now we have focused on the case of purely leptonic final states.
For the purely hadronic ones there are many more diagrams, as the
photon can interact with all four final-state fermions. In order to make
efficient use of the results presented in the previous sections, we first 
introduce some short-hand notations based on the results for the purely 
leptonic~($LL$) final states. These short-hand notations involve the summation 
of virtual and real corrections to the differential cross-section. 
For instance, the virtual corrections originating from the first diagram of 
Fig.~\ref{fig:1} can be combined with the corresponding real-photon correction 
into the contribution $d\sigma_{LL}^{(4)}(k_1;k_1^{\prime}|p_2)$.
In a similar way, virtual and real five-point corrections can be combined into
$d\sigma_{LL}^{(5)}(k_1;k_1^{\prime}|k_2;k_2^{\prime})$. The gauge-restoring
``Coulomb'' contribution will be indicated by 
$d\sigma^{\mbox{\scriptsize C}}(p_1|p_2)$. In terms of this notation the 
non-factorizable differential cross-section for purely leptonic final states 
becomes
\begin{equation}
\label{eq:LL}
  d\sigma_{LL}(k_1;k_1^{\prime}|k_2;k_2^{\prime}) = 
                    d\sigma_{LL}^{(4)}(k_1;k_1^{\prime}|p_2)
                  + d\sigma_{LL}^{(4)}(k_2;k_2^{\prime}|p_1)
                  + d\sigma_{LL}^{(5)}(k_1;k_1^{\prime}|k_2;k_2^{\prime})
                  + d\sigma^{\mbox{\scriptsize C}}(p_1|p_2).
\end{equation}
Analogously the non-factorizable differential cross-section for a purely 
hadronic final state~($HH$) can be written in the following way 
\begin{eqnarray}
\label{eq:HH}
  \lefteqn{\hspace*{-6mm}d\sigma_{HH}(k_1;k_1^{\prime}|k_2;k_2^{\prime}) = 
        3\times 3 \biggl[
        \frac{1}{3}\,d\sigma_{LL}^{(4)}(k_1;k_1^{\prime}|p_2)
      + \frac{2}{3}\,d\sigma_{LL}^{(4)}(k_1^{\prime};k_1|p_2)
      + \frac{1}{3}\,d\sigma_{LL}^{(4)}(k_2;k_2^{\prime}|p_1)}
                  \biggr. \nonumber \\[1mm]
              & & \biggl.
    {}+ \frac{2}{3\,}\,d\sigma_{LL}^{(4)}(k_2^{\prime};k_2|p_1)
      + \frac{1}{3}\cdot \frac{1}{3}\,
        d\sigma_{LL}^{(5)}(k_1;k_1^{\prime}|k_2;k_2^{\prime})
      + \frac{2}{3}\cdot \frac{1}{3}\,  
        d\sigma_{LL}^{(5)}(k_1^{\prime};k_1|k_2;k_2^{\prime})
                  \biggr. \nonumber \\[1mm]
              & & \biggl.
    {}+ \frac{1}{3}\cdot \frac{2}{3}\,   
        d\sigma_{LL}^{(5)}(k_1;k_1^{\prime}|k_2^{\prime};k_2)
      + \frac{2}{3}\cdot \frac{2}{3}\,   
        d\sigma_{LL}^{(5)}(k_1^{\prime};k_1|k_2^{\prime};k_2)
      + d\sigma^{\mbox{\scriptsize C}}(p_1|p_2)
  \biggr].
\end{eqnarray}
In order to keep the notation as uniform as possible, the momenta of the 
final-state quarks are defined along the lines of the purely leptonic case 
with $k_i$ ($k^{\prime}_i$) corresponding to down- (up-) type quarks. If one 
would like to take into account quark-mixing effects, it suffices to add the 
appropriate squared quark-mixing matrix elements ($|V_{ij}|^2$) to the overall 
factor. Note that top quarks do not contribute to the double-pole residues, 
since the on-shell decay $W \to t b$ is not allowed. Therefore the 
approximation of massless final-state fermions is still justified. 

For a semileptonic final state (say $HL$), when the $W^+$ decays 
hadronically and the $W^-$ leptonically, one can write
\begin{eqnarray}
\label{eq:HL}
    \lefteqn{\hspace*{-6mm} d\sigma_{HL}(k_1;k_1^{\prime}|k_2;k_2^{\prime}) =
             3 \biggl[ 
               \frac{1}{3}\,d\sigma_{LL}^{(4)}(k_1;k_1^{\prime}|p_2)
             + \frac{2}{3}\,d\sigma_{LL}^{(4)}(k_1^{\prime};k_1|p_2)
             +              d\sigma_{LL}^{(4)}(k_2;k_2^{\prime}|p_1)}
               \biggr. \nonumber \\
                                                  & &   \biggl. {}
      + \frac{1}{3}\,d\sigma_{LL}^{(5)}(k_1;k_1^{\prime}|k_2;k_2^{\prime})
      + \frac{2}{3}\,d\sigma_{LL}^{(5)}(k_1^{\prime};k_1|k_2;k_2^{\prime})
      + d\sigma^{\mbox{\scriptsize C}}(p_1|p_2)
                                                        \biggr].
\end{eqnarray}

Upon integration over the decay angles, the functions $d\sigma_{LL}^{(5)}$
and $d\sigma_{LL}^{(4)}$ become symmetric under 
$k_i \leftrightarrow k_i^{\prime}$. As a result, expressions 
(\ref{eq:HH}) and (\ref{eq:HL}) take on the form of (\ref{eq:LL}) multiplied
by the colour factors 9 and 3, respectively. These are precisely the colour
factors that also arise in the Born cross-section. Therefore, after
integration over the decay angles, the relative non-factorizable correction is
the same for all final states. This universality property holds for all 
situations that exhibit the $k_i \leftrightarrow k_i^{\prime}$ symmetry.
This means that there is only sensitivity to the charges of the 
decaying particles.

At this point one may wonder how non-factorizable corrections affect 
$Z$-pair-mediated and $ZH$-mediated four-fermion final states. In those cases, 
only five-point functions contribute, of which there are four contributions, 
as in Eq.\,(\ref{eq:HH}). However, in contrast to Eq.\,(\ref{eq:HH}), the 
charge factors are pair-wise opposite, such that integration over
the decay angles leads to a vanishing result. Thus $\cal{O}(\alpha)$
non-factorizable corrections to invariant-mass distributions
in $Z$-pair-mediated or $ZH$-mediated four-fermion processes vanish.
This can be viewed as a consequence of the zero charges of the 
decaying particles.

\section{Conclusions}
\label{conclusions}

In this paper we studied two methods to evaluate non-factorizable QED 
corrections
in the double-pole, soft-photon approximation. We derived results for $W$-pair 
production, which are valid a few widths above threshold.

One technique (DMI) is an extension of that of Ref.\,\cite{melyak} in the 
sense that the virtual and real photonic corrections are clearly separated
and also regularized by a photon mass $\lambda$ and charged-fermion masses
$m_{1}$ and $m_{2}$. The resulting formulae are rather complicated and are
different from those of \cite{melyak}.

The second method (MST) extends the standard technique in the sense that 
five-point
bremsstrahlung interference terms are decomposed into four-point terms and that
the soft-photon approximation is used from the start in the evaluation of real 
and virtual $n$-point functions. The results obtained with this method are 
much simpler 
than the DMI ones, but the two are in complete numerical agreement.

The MST can be easily generalized to more involved final states by a 
straightforward
extension of the decomposition of five-point functions to the decomposition 
of $n$-point functions.

The methods and most of the actual formulae in this paper can 
also be applied to $ZZ$, $ZH$ production
and to top-quark pair production with subsequent $Wb$ decays. In the latter 
case
the top-quark, $W$
and $b$ take the r\^ole of $W$, $\nu$ and $\ell$, the gluon that of the 
photon. In
that case  the DMI formulae are directly applicable since no assumption on the 
neutrino mass was made. The MST formulae would need a small modification.

\appendix

\section{Feynman-parameter integrals}

\subsection{The on-shell four-point function}
\label{app:feynman/D1234}
In this appendix we use Ref.\,\cite{denner} to present 
a compact expression 
for the on-shell four-point function $D_{1234}$, which appears in the 
decomposition 
of the virtual five-point function in Sect.\,\ref{sec:decomp/application}.
The result for $D_{1234}^{R}$ can be obtained from $D_{1234}$ by the  
substitutions $p_{1}\to-p_{1}$ and $k_{1}\to-k_{1}$. The results of \cite{gj}
provide an independent check on the formula presented here.

As was mentioned before, the four-point function $D_{1234}$ has to be 
calculated 
without soft-photon approximation and in the on-shell limit.
The resulting function, defined as
\be
          D_{1234} = \int
          \frac{d^{4} k}{(2 \pi)^{4}}
          \frac{1}
          {
           [(k - p_{1})^{2} - M_{W}^{2}] 
           [(k + p_{2})^{2} - M_{W}^{2}]
           [(k - k_{1})^{2} - m_{1}^{2}] 
           [(k + k_{2})^{2} - m_{2}^{2}]
           },
\end{equation}
does not contain any divergences.
Upon neglecting the fermion masses $m_{1}$ and $m_{2}$, the answer reads
$$
         M_{W}^{2} a (x_{1} - x_{2}) D_{1234}
         =
         \frac{i}{16\pi^{2}}
         \sum_{k=1}^{2}
         (-1)^{k}
         \Biggl\{
             \li\biggl(r_{14};-x_{k}\biggr)
            + \li\biggl(\frac{1}{r_{14}};-x_{k}\biggr)  
         \Biggr.
\ \ \ \ \ \ \ \ \ \ \ \ \ \ \ \ \ \ \ \ \ 
$$
\begin{equation}
         \Biggl.
          {}- \li\biggl(- \,
\frac{2(p_{2}\cdot k_{1})}{M_{W}^{2}} - io;-x_{k}\biggr)
            - \li\biggl(- \,
\frac{M_{W}^{2}}{2(p_{1}\cdot k_{2})} + io;-x_{k}\biggr)
            + \ln(-x_{k}) 
\ln\biggl[\frac{(p_{1}\cdot k_{2})}{(k_{1}\cdot k_{2})}\biggr]
         \Biggr\},
\end{equation}
where $r_{14}$ is a solution of the equation
\be
          r_{14} + \frac{1}{r_{14}} = - \,\frac{s - 2 M_{W}^{2}}
{M_{W}^{2}} - io.
\ee
The quantities $x_{1,2}$ are the solutions of the equation
\be
          a x^{2} + b x + c + i d o = 0,
\ee
with coefficients
$$
           a = 2(k_{1}\cdot k_{2}) - 2(p_{2}\cdot k_{1}), \ \ \ 
           c = 2(k_{1}\cdot k_{2}) - 2(p_{1}\cdot k_{2}), 
$$
\be
           b = M_{W}^{2} + 
           \frac{4(p_{1}\cdot k_{2}) (p_{2}\cdot k_{1})}{M_{W}^{2}} - 
           \frac{4(p_{1}\cdot p_{2}) (k_{1}\cdot k_{2})}{M_{W}^{2}},\ \ \ 
           d = - 2(k_{1}\cdot k_{2}).
\ee

\subsection{The infrared-finite four-point function}
\label{app:feynman/IRF}

In this appendix we briefly describe some of the details of the 
Feynman-parameter integral belonging to $D_{0123}$ 
(see Sect.\,\ref{sec:feynman/IRF}). The integral to be 
evaluated is given by Eq.\,(\ref{eq:feynman/IRF/appendix}). In the notation
adopted in Sect.\,\ref{sec:feynman} this integral reads
\be
  D_{0123} = -\frac{i}{4 \pi^{2} D_{2}} 
             \int\limits_{0}^{1} d\xi_{1} \int\limits_{0}^{1-\xi_{1}} 
             \frac{d\xi_{2}}{[y_{0} \xi_{1} + \xi_{2}] [p^2(\xi)-io]}, 
\ee
with
\be
  \frac{p^2(\xi)}{4 M_W^2} = \xi_{2}^{2} (1 - \zeta) 
         + \xi_{1} \xi_{2} (- 1 - \zeta + x_{s} + 1/x_{s})
         + \xi_{1} + \xi_{2} \zeta 
         + \frac{m_{1}^{2}}{M_{W}^{2}} (1 - \xi_{1} - \xi_{2})^{2}.
\ee
We perform the following change of variables:
\be
  \xi_{1} = \frac{1}{1 + t + u} \ \ \ \mbox{and}\ \ \ 
  \xi_{2} = \frac{t}{1 + t + u}.
\ee
Accordingly, the area of integration changes from
\be
  \xi_{1}>0, \ \ \ \xi_{2}>0, \ \ \ \xi_{1} + \xi_{2} < 1,
\ee
to 
\be
  0 < t < \infty, \ \ \ 0 < u < \infty.
\ee
The Jacobian of the transformation is given by
\be
  \biggl| \frac{\partial(\xi_{1}\xi_{2})}{\partial(t u)} \biggr|
   = \frac{1}{[1 + u + t]^{3}}.
\ee
The final integral to be evaluated looks like
\be
  M_{W}^{2} D_{0123} = -\,\frac{i}{16 \pi^{2} D_{2}} \int\limits_{0}^{\infty}
          \frac{du\,dt}{[y_{0} + t] [t^{2} + t (x_{s} + 1/x_{s}) + 1 + 
                        u (1 + t \zeta) + \frac{m_{1}^{2}}{M_{W}^{2}} u^{2}]}.
\ee
The second expression in the denominator of the integrand is linear in $u$ up
to the small term $u^2\,m_{1}^{2}/M_{W}^{2}$, which is needed to regularize 
the collinear divergences of the integral. When performing partial
fractioning of the integrand, this term should not be neglected; it has to be 
treated as a small parameter.

Performing first the integration over $u$ and then the integration over $t$, 
we obtain the final result (\ref{eq:feynman/IRF/D0123}) for $D_{0123}$,
expressed in terms of logarithms and dilogarithms.

\subsection{The infrared-divergent four-point function}
\label{app:feynman/IRD}

Next we present a few steps in the calculation of the Feynman-parameter
integral belonging to $D_{0134}$ (see Sect.\,\ref{sec:feynman/IRD}). The first
step involves the integration over momentum space, as represented by 
Eq.\,(\ref{eq:feynman/IRD/appendix}). The contour will be closed in the 
lower half of the complex $k_0$-plane, where only one pole is situated at 
$k_0=\sqrt{\vec{k}^2+\lambda^2-io}$. We introduce cylindric variables in the 
$\pvec(\xi)$ direction:
\be
  I_{0134}(\xi) = - \frac{i}{8 \pi^{2}} \int 
           \frac{\rho\, d\rho\, d z}{\sqrt{\rho^{2} + z^{2} + \lambda^{2}}\ 
           \Bigl[ - |E(\xi)| \, \sqrt{\rho^{2} + z^{2} + \lambda^{2}} 
                  - |\vec{\, p}(\xi)|\, z + A(\xi) + i o 
           \Bigr]^{3}}.
\ee
The integration over $\rho$ becomes trivial:
\be
\label{eq:app:feynman/IRD}
  I_{0134}(\xi) = \frac{i}{16 \pi^{2} |E(\xi)|} \,
\int \limits_{-\infty}^{\infty} 
          \frac{d z }{\Bigl[- |E(\xi)| \, \sqrt{z^{2} + \lambda^{2}} 
                            - |\vec{\, p}(\xi)|\, z + A(\xi) + i o \Bigr]^{2}}.
\ee
The last integration simplifies if one uses the representation:
\be
  I_{0134}(\xi) = \frac{i}{16 \pi^{2} \lambda \, |E(\xi)|} \,
           \frac{\partial}{\partial |E(\xi)|}\,
           \Biggl\{ \,\int \limits_{-\infty}^{\infty} 
                    \frac{d z }{\sqrt{z^{2} + 1}\,
            \Bigl[- |E(\xi)| \,\sqrt{z^{2} + 1} - |\vec{\, p}(\xi)|\, z 
                           + A(\xi)/\lambda + i o 
                    \Bigr]}
           \Biggr\}. 
\ee
Introducing the standard variable transformation $\ t=\sqrt{z^{2}+1}-z\,$,
the final integration can be performed, leading to the result given in 
Eq.\,(\ref{eq:feynman/IRD/1}).

The second stage of the calculation involves the integration over the 
Feynman parameters. We start with Eq.\,(\ref{eq:feynman/IRD/1}). This time  
we combine a change of variables, analogous to the one utilized in 
the previous appendix, with a rescaling :
\be
  \xi_{3} = \frac{t}{1 + u + t}, \ \ \ 
  \xi_{2} = \frac{1}{1 + u + t} \ \ \ \mbox{followed by} \ \ \ 
  t \to \frac{M_{W}}{m_{2}}\,t,  \ \ \ 
  u \to \frac{M_{W}}{m_{1}}\,u.
\ee
In this way $D_{0134}$ takes the form
\be
  D_{0134} = - \frac{i}{4 \pi^{2}}\,\frac{M_{W}^{2}}{m_{1} m_{2}}
      \int \limits _{0}^{\infty} d u \,d t \,
      \frac{\partial}{\partial p^{\prime\,2}}\,
      \biggl\{ \frac{1}{\sqrt{A^{\prime\,2} - \lambda^{2} p^{\prime\,2}}}
               \ln\biggl( 
            \frac{A^{\prime} - \sqrt{A^{\prime\,2}-\lambda^{2} p^{\prime\,2} }}
                 {A^{\prime} + \sqrt{A^{\prime\,2}-\lambda^{2} p^{\prime\,2} }}
                         \,\biggr)
             \biggr\},
\ee
where the definitions of $A^{\prime}$ and $p^{\prime\,2}$ have now changed. 
Those quantities are rescaled versions of $A$ and $p^2$. The rescaling is 
performed in such a way that $p^{\prime\,2}$ changes to
\be
  \frac{p^{\prime\,2}}{4 M_{W}^{2}} = t^{2} + u^{2} 
               - \frac{s_{12}}{m_{1} m_{2}}\, t u
               + \frac{\zeta^{\prime} M_{W}}{m_{2}}\, t + \frac{M_{W}}
{m_{1}}\, u + 1.
\ee
In order to linearize this expression with respect to $u$, one has to 
introduce one more variable transformation
\be
  t = t^{\prime} + c\, u, \ \ \ 
  \mbox{with} \ \ \ c = \frac{m_{1} m_{2}}{s_{12}}, 
\ee
which leads to 
\be
  \frac{p^{\prime\,2}}{4 M_{W}^{2}} = t^{\prime\,2} 
               - \frac{s_{12}}{m_{1} m_{2}}\, t^{\prime} u 
               + \frac{\zeta^{\prime} M_{W}}{m_{2}}\, t^{\prime}   
               + \frac{M_{W}}{m_{1}}\, u + 1.
\ee
After changing the order of integration according to  
\be
  \int\limits_{0}^{\infty}du \int\limits_{0}^{\infty}dt\, \to
  \int\limits_{0}^{\infty}du \int\limits_{-c\, u}^{\infty}dt^{\prime} 
  =
  \int\limits_{0}^{\infty}dt^{\prime} \int\limits_{0}^{\infty}du 
  +\int\limits_{-\infty}^{0}dt^{\prime} \int\limits_{-t^{\prime}/c}^{\infty}du,
\ee
one can perform the rest of the Feynman-parameter integrations to obtain the 
final result (\ref{eq:feynman/IRD/D0134}) for $D_{0134}$.

\section{Why \boldmath ${\cal R}$ vanishes}
\label{app:decomp}

\begin{figure}[t]
\unitlength 1cm
\begin{center}
\begin{picture}(17,8)
\put(7,4.5){\makebox[0pt][c]{$(k_0)$}}
\put(14,4.5){\makebox[0pt][c]{$(k_0)$}}
\put(5.5,-0.5){\makebox[0pt][c]{(a)}}
\put(11.5,-0.5){\makebox[0pt][c]{(b)}}
\put(1,-6){\includegraphics{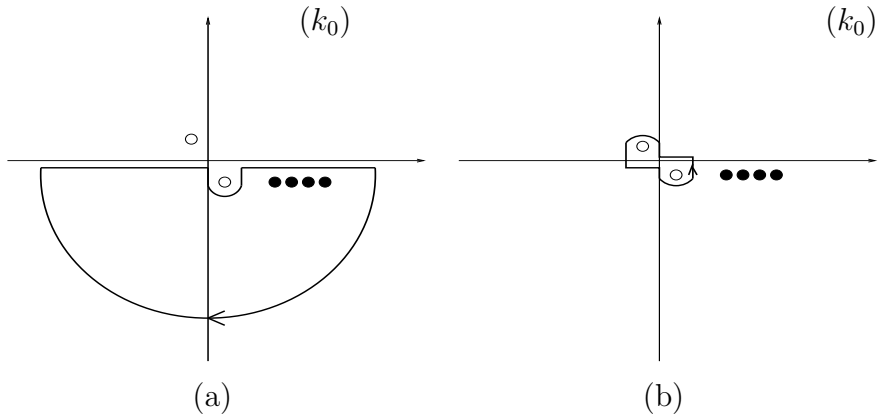}}
\end{picture}
\end{center}
\caption[]{The pole structure of ${\cal R}$ in the soft-photon approximation 
           (a) and the transformation of the integration contour in the complex
           $k_0$-plane (b). The solid circles indicate the particle poles, the
           open circles the photon poles.}
\label{fig:decomp2}
\end{figure} 

In this appendix it will be shown that the second term in 
Eq.\,(\ref{eq:decomp/real/2}), given by 
\be
  {\cal R} = \int\frac{d^{4}k}{(2\pi)^{4}N_{0}}\,\pole\,
             \frac{\sum r_{i}(k\cdot v_{i}) + 2 a \lambda^{2}}
             {N_{1}N_{2}N_{3}N_{4}},
\ee
is actually zero. In this integral the photon is not necessarily on-shell, 
because the residue is not taken in the photon pole. However, by power 
counting we can conclude that only soft photons give a noticeable contribution 
to the integral. All other contributions are formally of higher order in the 
expansion in powers of $\Gamma_W/M_W$. Therefore we use the soft-photon 
approximation to evaluate this integral. As a result, all particle poles are
situated in the same half-plane of the complex $k_0$ variable, as is shown in
Fig.~\ref{fig:decomp2}(a). 

Next one can deform the integration contour in the way depicted in 
Fig.~\ref{fig:decomp2}(b). Note that the orientation of the contour is 
reversed.
Figure~\ref{fig:decomp2} shows that the sum of the particle-pole residues is 
equal to the sum of the photon-pole residues with the opposite sign. This is 
a consequence of the soft-photon approximation and of the fact that all 
particle poles turned out to be in the same half-plane of the complex $k_0$
variable. The latter is the result of the transformation 
(\ref{eq:decomp/real/move_the_poles}) introduced in 
Sect.\,\ref{sec:decomp/real}.

Let us consider the following general integral
\be
  {\cal R}(p) = \int\frac{d^{4}k}{(2\pi)^{4}N_{0}}\,\pole\,
                \frac{p\cdot k}{N_{1}N_{2}N_{3}N_{4}},
\ee
where $p^{\mu}$ is an arbitrary vector. In the soft-photon approximation the 
denominators can be written as $N_0 = k^2 - \lambda^2 + io\,$ and 
$\,N_{i} = 2(p_{i}\cdot k) + p_i^2 - m_i^2 +io$. As mentioned
in Sect.\,\ref{sec:decomp/real}, the momenta $p_i$ are time-like and have 
positive energy components, i.e.~$E_i \ge |\pvec_i|$. For simplicity we take 
the photon to be massless, i.e.~$\lambda^{2}=0$, but the arguments that follow 
do not depend on this. Deforming the integration contour as described above, 
and subsequently taking the residues in the photon poles, one can write
\be
  {\cal R}(p) = 2\pi i \int \frac{d^{3}k}{(2\pi)^{4} \,2 \omega}\,
                \frac{(p\cdot k)}{N_{1}N_{2}N_{3}N_{4}}
                \Biggl.\Biggr|_{\omega=-|\vec{k}|+io}    
                +
                2\pi i \int \frac{d^{3}k}{(2\pi)^{4} \,2 \omega}\,
                \frac{(p\cdot k)}{N_{1}N_{2}N_{3}N_{4}}
                \Biggl.\Biggr|_{\omega=|\vec{k}|-io}.
\ee
In spherical coordinates this takes the form
\begin{eqnarray}
  {\cal R}(p) &=& i\pi \int\limits_{0}^{\infty}
        \frac{d|\vec{k}|\, d^{2}\Omega_k}{(2\pi)^{4}}\,
        \frac{|\vec{k}|^{2}\,(E  + f(\Omega_k) |\pvec|)}{\prod\limits_{i=1}^4
              \Bigl[ - 2 |\vec{k}|\,(E_{i} + f_{i}(\Omega_k) |\pvec_i|) 
                     + p_i^2 - m_i^2 +io
              \Bigr]} \nonumber \\[1mm]
              &\,\,\,+\!\!\!& i\pi \int\limits_{0}^{\infty}
        \frac{d|\vec{k}|\, d^{2}\Omega_k}{(2\pi)^{4}}\,
        \frac{|\vec{k}|^{2}\,(E - f(\Omega_k) |\pvec|)}{\prod\limits_{i=1}^4
              \Bigl[ 2 |\vec{k}|\,(E_{i} - f_{i}(\Omega_k) |\pvec_i|) 
                    + p_i^2 - m_i^2 +io
              \Bigr]}.
\end{eqnarray}
In the second term one can make a change of variables according to
$|\vec{k}|\to -|\vec{k}|$ and $\vec{n}_{k}\to -\vec{n}_{k}$, to obtain
\be
  {\cal R}(p) = i\pi \int\limits_{-\infty}^{\infty}
        \frac{d|\vec{k}|\, d^{2}\Omega_k}{(2\pi)^{4}}\,
        \frac{|\vec{k}|^{2}\,(E + f(\Omega_k) |\pvec|)}{\prod\limits_{i=1}^4 
              \Bigl[ - 2 |\vec{k}|\,(E_{i} + f_{i}(\Omega_k) |\pvec_i|) 
                     + p_i^2 - m_i^2 +io
              \Bigr]}.
\ee
This integral is ultraviolet-finite and all poles are situated in the same 
half-plane of the complex variable $|\vec{k}|$, since $E_i \ge |\pvec_i|\,$
and $\,|f_{i}(\Omega_k)| \le 1$.
By closing the contour in the opposite half-plane, one 
finds ${\cal R}(p) = 0$. From this it trivially follows that ${\cal R} = 0$.

\section{Special functions and integrals in the DMI method}

\subsection{The functions $F_{1}$ and $F_{2}$}
\label{app:dmi/f1_f2}

In this appendix we present the functions $F_1$ and $F_2$, which are used in 
the calculations in Sect.\,\ref{sec:dmi}. The function $F_1$ is defined as 
\be
  F_{1}(a;\beta | x_{i}) = \int\limits_{-1}^{1}\frac{d x}{x - a}\,
          \ln(1+\beta x)\,\biggl[ \theta(x-x_{i}) - \theta(x_{i} - x) \biggr].
\ee
Here $a$ is a complex number with a non-zero imaginary part, and 
$\beta$ and $x_{i}$ are real numbers with absolute value smaller than $1$. 
The analytical expression for this function is
given by
\begin{eqnarray}
  F_{1}(a;\beta | x_{i}) &=&
      - 2\,\Li\biggl( \frac{1 + x_{i}\beta}{1 + a \beta} \biggr)
      + \Li\biggl( \frac{1 - \beta}{1 + a \beta} \biggr)
      + \Li\biggl( \frac{1 + \beta}{1 + a \beta} \biggr)
      + \ln\biggl( \frac{\beta (1 + a)}{1 + a \beta} \biggr)\ln(1 - \beta)
      \nonumber \\[2mm]
                         & &
      {}+ \ln\biggl( \frac{\beta (a - 1)}{1 + a \beta} \biggr)\ln(1 + \beta)
      - 2\ln\biggl(\frac{\beta (a-x_{i})}{1+a\beta} \biggr)\ln(1 + x_{i}\beta).
\end{eqnarray}
In addition we need this function in the special case $a = x_{i}$, without 
non-zero imaginary part. There, the integral $F_{1}(x_{i};\beta|x_{i})$ 
is logarithmically divergent. This is a collinear divergence and should be
regularized by keeping the small non-zero fermion masses.
The answer in this case is
\be
  F_{1}(x_{i};\beta | x_{i}) =
      \ln(1 + x_{i}\beta) \ln\biggl( \frac{4 E_{i}^{2}}{m_{i}^{2}} \biggr)
      - \Li\biggl( \beta\frac{1 + x_{i}}{1 + x_{i}\beta} \biggr) 
      - \Li\biggl( \beta\frac{x_{i} - 1}{1 + x_{i}\beta} \biggr).
\ee  

The other function, $F_2$, is defined as
\be
  F_{2}(a|x_{i}) = \int\limits_{-1}^{1}\frac{d x}{x - a}\,
                   \biggl[ \theta(x-x_{i}) - \theta(x_{i} - x) \biggr].
\ee
For $a$ and $x_i$ the same restrictions as indicated for the function 
$F_1$ apply. The corresponding analytical expressions are
\be
  F_{2}(a|x_{i}) = - 2 \ln(x_{i} - a) + \ln(- 1 - a) + \ln(1 - a),
\ee
and
\be
  F_{2}(x_{i}|x_{i}) = \ln\biggl( \frac{4 E_{i}^{2}}{m_{i}^{2}} \biggr). 
\ee

\subsection{The azimuthal principal-value integral}
\label{app:dmi/azimuthal}

In this appendix we present the result for the azimuthal principal-value 
integral, used in Sect.\,\ref{sec:dmi}:
\be
  I_{\phi} = {\cal P}\;\Biggl( \int\limits_{0}^{2\pi} \frac{d\phi}{2\pi}\,
             \frac{1}{A - B \cos\phi} \Biggr),
\ee
with
\be
  A = ( v_{1} - v_{2}\cos\theta_{12})\,\cos\theta \ \ \ \ \mbox{and} \ \ \ \
  B = v_{2}\sin\theta_{12}\sin\theta.
\ee
The principal-value integration yields
\be
  I_{\phi} = \left\{ \begin{array}{l} 
                        + \frac{1}{\sqrt{A^{2}-B^{2}}}\ \ \ 
                        \mbox{for}\ \ \ A/B\in(+1,+\infty)\ \ \
                        \mbox{or equivalently}\ \ \ \cos\theta\in(+x_{+},+1) 
                        \\
                        \\
                        -\frac{1}{\sqrt{A^{2}-B^{2}}}\ \ \
                        \mbox{for}\ \ \ A/B\in(-\infty,-1)\ \ \
                        \mbox{or equivalently}\ \ \ \cos\theta\in(-1,-x_{+})
                     \end{array},
             \right.
\ee
where 
\be
  x_{+} = \sqrt{ \frac{v_{2}^{2}(1-x_{12}^{2})}
                      {v_{1}^{2} + v_{2}^{2} - 2v_{1}v_{2}x_{12}} }\ \ \ \
  \mbox{and}\ \ \ \
  \sqrt{A^{2} - B^{2}} = |\vec{v}_{1} - \vec{v}_{2}|\,\sqrt{x^{2} - x_{+}^{2}}.
\ee
In non-collinear situations one can take $v_{1,2}\to 1$, resulting in 
$x_+ = \sqrt{(1 + x_{12})/2}\,$.

\subsection{The functions $\ff_{1}$, $\ff_{2}$ and ${\cal K}$}
\label{app:dmi/ff1_ff2}

In this appendix we present the functions $\ff_{1}$, $\ff_{2}$, and ${\cal K}$,
used in Sect.\,\ref{sec:dmi} for the infrared-divergent four- and five-point 
functions. 

The function $\ff_{1}$ is defined as 
\be
  \ff_{1}(x_{+};v) = \int\limits_{x_{+}}^{1}
                     \frac{d x}{\sqrt{x^{2} - x_{+}^{2}}}\,\frac{1}{1 - v x}\,
                     \ln\biggl( \frac{1-v x}{1+v x} \biggr).
\ee
Here, $x_+$ is real with $0 \le x_+ < 1$, and the quantity $v$ is real and 
close to unity. For $v\to 1$ the answer for this integral is given by
\be
  \ff_{1}(x_{+};v) = \frac{1}{\sqrt{1-x_+^2}}\,
       \Biggl[
       - \frac{1}{2}\ln^{2}\biggl( \frac{1-v}{2} \biggr) - \frac{\pi^{2}}{6}
       + \frac{1}{2}\ln^{2}(1-x_{+}^{2}) - \ln(x_{+})\,\ln(1-x_{+}^{2})
       \Biggr].
\ee 

The function $\ff_{2}$ is defined as
\be
  \ff_{2}(x_{+}) =  \int\limits_{x_{+}}^{1}
                    \frac{d x}{\sqrt{x^{2} - x_{+}^{2}}}\,\frac{1}{1 + x}\,
                    \ln\biggl( \frac{1 - x}{1 + x} \biggr),
\ee
which amounts to
\be
  \ff_{2}(x_{+}) = \frac{1}{\sqrt{1-x_+^2}}\,
                   \Biggl[ \Li(x_{+}^{2}) - \frac{\pi^{2}}{6}
                           + \ln(x_{+})\,\ln(1 - x_{+}^{2})
                   \Biggr].
\ee
In our explicit formulae, the functions $\ff_{1}$ and $\ff_{2}$ always enter 
as a sum. This sum can be represented in a compact form:
\be
  \ff_{1}(x_{+};v) + \ff_{2}(x_{+}) = \frac{1}{\sqrt{1-x_+^2}}\,
        \Biggl[
           - \frac{1}{2}\ln^{2}\biggl( \frac{1 - v}{2} \biggr)
           - \Li\biggl( \frac{x_{+}^{2}}{x_{+}^{2}-1} \biggr)
           - \frac{\pi^{2}}{3}
        \Biggr].
\ee

The function ${\cal K}$ is defined as
\be
  {\cal K}(A;B|x_{0};\mu^{2}) = \int\limits_{-1}^{1}
          \frac{d x}{(x + A) \sqrt{(x - x_{0})^{2} + \mu^{2}}}\,\ln(B - x),
\ee
$A$ being a complex number with a non-zero imaginary part, and 
$B$ being real and larger than $1$. The quantities $x_0$ and $\mu$ are real,
with $|x_0|<1$ and $\mu^2 \ll 1$.
The resulting analytical expression is somewhat more complicated:
\begin{eqnarray}
  {\cal K}(A;B|x_{0};\mu^{2}) \!\!\!&=&\!\!\! \frac{-1}{A + x_{0}}
         \Biggl\{ \li\biggl( 1;\frac{A+1}{A+x_{0}} \biggr)
                  - \li\biggl( 1;\frac{A+x_{0}}{A-1} \biggr)
                  - \li\biggl( \frac{B-x_{0}}{B-1};\frac{A+1}{A+x_{0}} \biggr)
         \Biggr. \nonumber \\[1mm]
                              \!\!\!& &\!\!\!
         \Biggl. \hphantom{\frac{-1}{A + x_{0}}a}
                {}+ \li\biggl( \frac{B+1}{B-x_{0}};\frac{A+x_{0}}{A-1} \biggr)
                  - \frac{1}{2}\ln^{2}\biggl( \frac{B-x_{0}}{B-1} \biggr)
                  + \ln(B-1)\,\ln\biggl( \frac{A+1}{A+x_{0}} \biggr)
         \Biggr. \nonumber \\[1mm]
                              \!\!\!& &\!\!\!
         \Biggl.\hphantom{\frac{-1}{A + x_{0}}a}
                {}+ \ln(B-x_{0})\ln\biggl( \frac{\mu^{2}}{4(1 - x_{0}^{2})}
                                           \frac{A-1}{A+x_{0}} \biggr)
         \Biggr\}.
\end{eqnarray}

\end{document}